%% file: main.tex
\documentclass[manuscript, screen]{acmart}

\AtBeginDocument{%
  }

\usepackage{booktabs}
\usepackage{multirow}
\usepackage{xcolor}
\usepackage{graphicx}
\usepackage{subcaption}
\usepackage{balance}
\usepackage{url}
\usepackage{tcolorbox}
\usepackage{tabularx}
\usepackage{xspace}
\usepackage{tabularray}
\usepackage{pgfplots}
\usepackage{pgfplotstable}
\usepackage{adjustbox}
\usepackage{colortbl}
\usepackage{algorithm}
\usepackage{algorithmic}
\usepackage{amsfonts}
\usepackage{longtable}
\usepackage{comment}
\usepackage{float}
\usepackage{hyperref}
\usepackage{listings}
\usepackage{xcolor}

\lstdefinelanguage{diff}{
  morecomment=[f][\color{red}]{-},
  morecomment=[f][\color{green!60!black}]{+},
  morecomment=[f][\color{gray}]{@},
}

\input{macros}

\title{On the Role of Fault Localization Context for LLM-Based Program Repair}

\author{Melika Sepidband}
\email{melikasp@yorku.ca}
\affiliation{%
  \institution{York University}
  \city{Toronto}
  \country{Canada}
}
\author{Hung Viet Pham}
\email{hvpham@yorku.ca}
\affiliation{%
  \institution{York University}
  \city{Toronto}
  \country{Canada}
}
\author{Hadi Hemmati}
\email{hemmati@yorku.ca}
\affiliation{%
  \institution{York University}
  \city{Toronto}
  \country{Canada}
}

\begin{document}

\begin{abstract}
    \input{Sections/0_Abstract}
\end{abstract}



\maketitle

\input{Sections/1_Introduction}

\input{Sections/2_Background}
\input{Sections/3_StudyDesign}
\input{Sections/4_Results}
\input{Sections/5_Qualitative_analysis}

\input{Sections/6_Discussion}
\input{Sections/7_Relatedwork}
\input{Sections/8_Threats_to_validity}
\input{Sections/9_Conclusion}

\balance

\bibliographystyle{ACM-Reference-Format}

\bibliography{references}
\end{document}

%% file: macros.tex
\usepackage{soul}
\usepackage{hyperref}
\usepackage{multirow}
\usepackage{listings}
\usepackage{fancybox}
\usepackage{enumitem}
\usepackage{graphicx}
\usepackage{color}
\usepackage{array}
\usepackage{amsmath}
\usepackage{xspace}
\usepackage{url}
\usepackage{tikz}
\usepackage{caption}
\usepackage{pgfplots}
\usepackage{balance}
\usepackage{subcaption}
\pgfplotsset{compat=1.16}
\usepackage{pgf-pie}
\usetikzlibrary{pgfplots.statistics,calc}
\usepackage{algorithm}
\usepackage{algorithmic}
\usepackage{adjustbox}
\usepackage{makecell}

\usepackage{tikz}
\usepackage{calc}

\usepackage{tcolorbox}
\makeatletter
\newcommand{\mybox}[1]{%
	\setbox0=\hbox{#1}%
	\setlength{\@tempdima}{\dimexpr\wd0+13pt}%
	\begin{tcolorbox}[boxrule=0.5pt, colback=white, arc=4pt,
		left=6pt,right=6pt,top=6pt,bottom=6pt,boxsep=0pt]
		#1
	\end{tcolorbox}
}

\definecolor{songcolor}{RGB}{191,191,191}



%% file: Sections/0_Abstract.tex
Fault Localization (FL) is a key component of Large Language Model (LLM)-based Automated Program Repair (APR), yet its impact remains underexplored. In particular, it is unclear how much localization is needed, whether additional context beyond the predicted buggy location is beneficial, and how such context should be retrieved. We conduct a large-scale empirical study on 500 SWE-bench Verified instances using GPT-5-mini, evaluating 61 configurations that vary file-, element-, and line-level context. Our results show that more context does not consistently improve repair performance. File-level localization is the dominant factor, yielding a 15–17× improvement over a no-file baseline. Expanding file context is often associated with improved performance, with successful repairs most commonly observed in configurations with approximately 6–10 relevant files. Element-level context expansion provides conditional gains that depend strongly on the file context quality, while line-level context expansion frequently degrades performance due to noise amplification. LLM-based retrieval generally outperforms structural heuristics while using fewer files and tokens. Overall, the most effective FL context strategy typically combines a broad semantic understanding at higher abstraction levels with precise line-level localization. These findings challenge our assumption that increasing the localization context uniformly improves APR, and provide practical guidance for designing LLM-based FL strategies.

%% file: Sections/1_Introduction.tex
\section{Introduction}
\label{sec:introduction}

Automated Program Repair (APR) has long promised to reduce the burden of software maintenance by automatically generating patches for buggy code~\cite{gazzola2018automatic,monperrus2018automatic}. The emergence of Large Language Models (LLMs) has dramatically advanced the state of the art, with models like GPT, Gemini, and Claude demonstrating remarkable ability to understand bug reports and generate correct fixes~\cite{xia2023automated,fan2023automated}. Despite these advances, successful repair still fundamentally depends on how effectively an LLM identifies where to focus its analysis within large and complex codebases.

A critical component of APR is Fault Localization (FL), which identifies the code locations most likely responsible for a bug~\cite{wong2016survey}. Traditional APR systems rely heavily on precise FL, assuming that narrowing the search space improves repair accuracy~\cite{liu2019you}. However, this assumption becomes less clear in the era of LLM-based repair. LLMs operate with large context windows capable of processing entire files or even multiple modules simultaneously, and may benefit from broader contextual information that captures semantic intent rather than only syntactic proximity to the bug.

This shift introduces a fundamental tension in LLM-based APR design. On one hand, providing more context gives the LLM additional information to understand the bug and its fix. On the other hand, excessive context may overwhelm the model, introduce irrelevant or distracting code, and dilute the signal of where the bug actually lies. While prior work has explored improving localization accuracy, it has not systematically examined how the type of localization context across multiple granularities (files, elements, and lines) affects repair success.

Early empirical evidence further challenges the assumption that increasingly precise localization necessarily leads to better repair outcomes. Recent work on LLM-based fault localization leverages the reasoning capabilities of LLMs to identify bug-relevant code locations~\cite{sepidband2026rgfl}. In this work, an ablation study on unresolved instances revealed an intriguing question: when the repair model fails, is it because fault localization was imprecise, or because the bug is inherently difficult regardless of localization quality? Preliminary analysis showed that even with perfect (ground-truth) localization, many instances remained unresolved, suggesting that localization precision alone does not determine repair success. This observation motivates the present study: a comprehensive investigation across \emph{all} instances (not just failures), examining all levels of context granularity to understand how much FL context actually helps and when it stops helping or even hurts.

Existing studies of LLM-based APR typically adopt a fixed localization strategy without systematically examining how different context levels influence repair outcomes~\cite{xia2023automated,zhang2024autocoderover}. While some work has compared FL techniques~\cite{koyuncu2019ifixr}, none has conducted a factorial study examining the interaction between file-level, element-level (functions/classes/global variables), and line-level context. As a result, a critical question remains underexplored: \emph{how much fault-localization context should we provide to maximize repair success?} Answering this question is essential for designing effective and cost-efficient APR pipelines, determining how effort should be allocated between localization and repair components.

In this paper, we present a large-scale empirical study systematically examining how FL context granularity affects LLM-based program repair. We design a factorial experiment with three orthogonal context dimensions:

\begin{itemize}
    \item \textbf{File-level context}: No files, buggy files, rule-based relevant files (based on files imported by and importing the buggy files), or LLM-retrieved relevant files
    \item \textbf{Element-level context}: No elements, buggy elements, Call Graph relevant elements, or LLM-retrieved relevant elements  
    \item \textbf{Line-level context}: No lines, buggy lines, context window ($\pm$10 lines), static code slicing, or LLM-retrieved lines
\end{itemize}

To study the impact of localization context without confounding factors, we assume a perfect localization by using the ground-truth (GT) patches in SWE-bench Verified to extract the buggy files, elements, and lines. This allows us to evaluate whether an LLM can successfully repair a bug when given the correct locations, and to systematically analyze how "expanding context" beyond these GT locations affects repair performance. We evaluate all 61 feasible configurations (combinations of different granularity) on 500 instances from SWE-bench Verified~\cite{jimenez2023swe}, a widely used benchmark for evaluating LLM-based APR, using GPT-5-mini, recording which instances each configuration resolves. This design enables us to analyze not only the individual effects of each context dimension but also their interactions, revealing when additional context helps, saturates, or harms repair performance.

\subsection{Findings}
We summarize the key insights from our empirical study on SWE-bench Verified. Across all configurations, consistent patterns emerge regarding how context granularity and expansion strategies influence repair success. To better understand the role of context, we formulate hypotheses and evaluate them using the Wilcoxon signed-rank test, allowing us to determine when expanded context significantly improves or degrades repair performance. The following findings highlight the dominant factors that influence repair success and provide practical guidance for designing effective FL pipelines for LLM-based program repair.

\begin{itemize}
    \item \textbf{File-level expansion consistently improves repair performance.}
    Introducing \emph{file} context provides the largest improvement over the no-context baseline, and expanding the context beyond \emph{"buggy files"} generally yields additional gains. Both \emph{"rule-based"} and \emph{"LLM-based"} file expansion significantly outperform buggy-only context.
    
    \item \textbf{LLM-based file retrieval outperforms rule-based expansion.}
    \emph{"LLM-retrieved files"} outperform \emph{"rule-based files"} in most configurations, and the improvement is statistically significant. The advantage is strongest when \emph{line} localization is precise (\emph{"buggy lines"})
    
    \item \textbf{LLM-based file retrieval incurs lower costs than rule-based expansion.}
    The histograms for the distribution of file counts and token budgets, along with the box plots, show that \emph{"LLM-based"} retrieval produces smaller and more consistent context sizes, while \emph{"rule-based"} methods lead to higher and more variable costs with frequent high-cost outliers. These differences are statistically significant, indicating that LLM approaches consistently require less context.
    
    \item \textbf{Element-level context provides conditional benefits, but expansion beyond buggy shows limited advantage.}
    Compared to providing no \emph{element} information, adding functions or classes generally improves repair outcomes. However, expanding beyond \emph{"buggy elements"} produces mixed results.\emph{"LLM-based element retrieval"} outperforms \emph{"buggy elements"} in most configurations, whereas \emph{"Call Graph"} expansion underperforms and does not achieve statistically significant gains over \emph{"buggy elements"}.

    \item \textbf{LLM-based element retrieval outperforms Call Graph retrieval.}
    \emph{"LLM-retrieved elements"} achieve better results than \emph{"Call Graph"} method in most configurations, particularly when paired with \emph{"LLM-retrieved files"}. The difference is statistically significant and consistently favors \emph{"LLM-based retrieval"}, suggesting that semantic retrieval yields higher-quality context than purely structural approaches.

    \item \textbf{Line-level expansion rarely improves repair performance and often introduces noise.}
    In contrast to higher granularities, expanded \emph{line}-level context generally does not outperform either precise \emph{"buggy line"} localization or even configurations without \emph{line} localization. \emph{"Context windows"} and \emph{"static code slicing"} frequently degrade performance and are harmful in many configurations. These results indicate that \emph{line}-level localization primarily serves to identify the editing location for LLM, rather than providing context to understand the bug better, and additional surrounding lines often dilute repair signals.
     
    \item \textbf{LLM-based line retrieval is a more effective expansion strategy compared to Context window and Code slicing expansions.}
    Although configurations that use \emph{"LLM-retrieved lines"} do not consistently exceed \emph{"buggy line"} or \emph{"no line"} settings, they significantly outperform configurations that use other expansion strategies. Therefore, when expanding beyond \emph{"buggy lines"}, \emph{"LLM-based line retrieval"} provides the most reliable gains.
    
    \item \textbf{Different granularities benefit from complementary localization strategies.}
    The best performance comes from combining \emph{LLM-based file} and \emph{element} context with precise \emph{"buggy lines"}, showing that broader context helps at higher levels, while \emph{line}-level localization should remain precise.

    \item \textbf{Combining complementary localization strategies offers further potential gains.}
    Different configurations resolve different subsets of instances, indicating that no single strategy is universally optimal. To better understand these differences, we conduct a qualitative analysis of cases where additional context helps or hurts performance, identifying categories and representative examples for both. Our findings show that some bugs benefit from broader context, while others require more focused localization, suggesting that adaptive or ensemble context strategies, which dynamically adjust localization based on bug characteristics, may outperform any fixed configuration.
    
\end{itemize}

%% file: Sections/2_Background.tex
\section{Background}
\label{sec:background}

\subsection{Automated Program Repair}

APR aims to automatically generate patches that make a buggy program satisfy its specification, typically operationalized as passing a test suite~\cite{gazzola2018automatic,monperrus2018automatic}. APR systems typically follow a pipeline consisting of fault localization, patch generation, and validation.

Early APR systems were predominantly search-based~\cite{weimer2009automatically}. GenProg~\cite{le2011genprog} applies genetic programming to explore candidate edits guided by suspiciousness rankings. Other search-based systems, such as RSRepair~\cite{qi2014strength} and SPR~\cite{long2015staged}, similarly rely on prioritized mutation spaces to control combinatorial explosion.

Semantics-based systems such as SemFix~\cite{nguyen2013semfix} and Angelix~\cite{mechtaev2016angelix} synthesize repairs using symbolic execution and constraint solving. Because symbolic reasoning is computationally expensive, these systems restrict synthesis to suspicious regions identified by FL, making localization quality directly impact repair feasibility.

More recently, learning-based and neural approaches have emerged, including sequence-to-sequence and transformer-based repair models~\cite{tufano2019empirical,mesbah2019deepdelta}. These systems still depend on localized input regions to constrain generation.

Because the space of possible edits in real-world software is extremely large, most APR techniques rely on FL to reduce, prioritize, or constrain candidate modification locations.

\subsection{Fault Localization}

Fault localization estimates the code locations responsible for failures. 
Most classical approaches are \textbf{spectrum-based fault localization (SBFL)}, which ranks program entities using test coverage statistics.

Representative suspiciousness metrics include Tarantula~\cite{jones2002visualization}, Ochiai~\cite{abreu2007accuracy}, and DStar~\cite{wong2013dstar}. SBFL has been extensively studied and benchmarked~\cite{wong2016survey}, and is widely integrated into APR pipelines. Beyond SBFL, several alternative localization strategies have been explored.

\textbf{Dynamic slicing} traces data- and control-dependencies from failing executions to reduce candidate regions~\cite{guo2018empirical,zhang2007study,mao2014slice}. While slicing can shrink the search space, empirical results show mixed impact on final repair success.

\textbf{Static structural analyses}, including call-graph and dependency-based methods, identify related functions or files through structural relationships~\cite{gong2014locating}. Static and hybrid structural heuristics leverage program call graphs and dependency relations to expand beyond individual suspicious statements.

\textbf{Information retrieval (IR)-based localization} links bug reports to source files via textual similarity~\cite{zhou2012should,li2020irbfl}.
IR-based file localization is particularly important for large-scale repository settings where statement-level ranking is impractical.

\subsection{LLM-Based and Agentic Repair}

The emergence of LLMs has fundamentally reshaped APR workflows. Recent systems increasingly adopt agent-based architectures that tightly couple LLMs with repository navigation, editing, and execution tools. For example, SWE-agent~\cite{xia2023automated} equips LLMs with a structured agent–computer interface for iterative code exploration and modification; AutoCodeRover~\cite{zhang2024autocoderover} employs a two-stage pipeline in which the model first navigates the repository using program-structure-aware search APIs before generating candidate patches; and Agentless~\cite{xia2024agentless} decomposes repair into localization, repair, and validation phases, performing hierarchical localization (file $\rightarrow$ function $\rightarrow$ edit location) via a combination of IR-based retrieval and LLM prompting. Furthermore, OpenHands~\cite{wang2024openhands} provides rich tooling and execution capabilities, prompting the model with substantial portions of the repository and issue context to directly propose fixes. Moatless Tools~\cite{antoniades2024swe} integrates Monte Carlo Tree Search with an LLM-based value function to support iterative exploration, backtracking, and multi-agent debate for patch selection. LocAgent~\cite{chen2025locagent} learns representations of code structure and execution behavior to improve localization accuracy, while Trae~\cite{gao2025trae} emphasizes aggressive tool-based exploration, enabling file inspection, symbol search, and test execution prior to patch generation. 

Rather than explicitly ranking suspicious statements as in traditional fault localization, these systems typically perform localization implicitly through repository navigation, heuristic retrieval, structural search, or LLM-guided reasoning over contextualized code. Although architecturally different from classical SBFL-based pipelines, these systems remain fundamentally dependent on localization decisions: the specific files, elements, and lines supplied as context strongly shape the generated patch. Thus, across both classical and LLM-based APR paradigms, FL serves as the mechanism that controls the search space and guides repair reasoning. The key difference lies not in whether localization matters, but in how localization information is selected, expanded, and incorporated into the repair process.

%% file: Sections/3_StudyDesign.tex
\section{Study Design}
\label{sec:study-design}

We design a factorial empirical study to systematically investigate how fault localization context influences LLM-based program repair. Our goal is to understand not only whether localization improves repair performance, but also how different context expansion strategies behave across multiple levels of abstraction. To this end, we study localization decisions at three complementary granularities (\emph{files}, \emph{elements}, and \emph{lines}). We organize the study around four research questions: three corresponding to these granularities and one examining their interactions.

\paragraph{\textbf{RQ1: What is the impact of file-level context on the APR's success?}\\
\textbf{RQ1.1: Does additional file-level context, beyond the precise buggy file location(s), help LLM-based APR?} We compare \emph{"rule-based"} and \emph{"LLM-retrieved relevant files"} (expanded context) against only \emph{"buggy files"} in the context.\\
\textbf{RQ1.2: Which retrieval strategy works better for file-level FL context expansion?} We compare \emph{"rule-based"} vs.\ \emph{"LLM-based"} relevant file retrieval. \\
\textbf{RQ1.3: What is the cost of rule-based versus LLM-based file-level context expansion?} We compare the cost of \emph{"rule-based"} and \emph{"LLM-based"} file-level expansion in terms of the number of retrieved files and total token budget across all instances, using distribution plots and statistical tests.}

\paragraph{\textbf{RQ2: What is the impact of element-level context on the APR's success?}\\
\textbf{RQ2.1: Does additional element-level context, beyond the precise buggy element location(s), help LLM-based APR?} We compare \emph{"Call Graph"} and \emph{"LLM-retrieved relevant elements"} against only \emph{"buggy elements"} in the context. \\
\textbf{RQ2.2: Which retrieval strategy works better for element-level FL context expansion?} We compare \emph{"Call Graph"} vs.\ \emph{"LLM-based"} relevant element selection.}

\paragraph{\textbf{RQ3: What is the impact of line-Level context on the APR's success?}\\
\textbf{RQ3.1: Does additional line-level context, beyond the precise buggy line location(s), help LLM-based APR?} We compare \emph{"context window"}, \emph{"code slicing"}, and \emph{"LLM-retrieved lines"} against only \emph{"buggy lines"} in the context.\\
\textbf{RQ3.2: Which retrieval strategy works better for line-level FL context expansion?} We compare \emph{"context window"} vs.\ \emph{"code slicing"} vs.\ \emph{"LLM-retrieved lines".}}

\paragraph{\textbf{RQ4: What are the characteristics of the best- and worst-performing FL context expansion strategies?}} We conducted a cross-dimensional analysis to examine how \emph{file}-, \emph{element}-, and \emph{line}-level contexts interact and which combinations of context expansion strategies yield the strongest and weakest overall repair performance.

\subsection{Dataset}

We evaluate our study using SWE-bench Verified~\cite{jimenez2023swe}, one of the most widely used benchmarks for evaluating APR systems in both academia and industry. It is a curated benchmark of 500 real-world software engineering instances drawn from widely used Python repositories. Each instance includes a GitHub issue describing the bug or feature request, the repository state before the fix, a developer-provided patch, and a test patch used to validate correctness.

The benchmark consists of 12 repositories, including Django, Matplotlib, Scikit-learn, and SymPy. Bugs range from small single-line edits to complex multi-file modifications involving substantial program reasoning. Its diversity and real-world provenance make SWE-bench Verified particularly well suited for studying how localization context influences repair performance.

\subsection{Model and Prompt}
\subsubsection{Model}
We use GPT-5-mini as the repair model. At the time this study was conducted, GPT-5-mini was the highest-performing small model on the SWE-bench leaderboard. This model provides a practical balance between reasoning capability and computational cost, enabling large-scale experimentation across many configurations. Each benchmark instance receives a single repair attempt per configuration to ensure consistent comparison. We opt to use smaller models (distilled models) due to the massive computation cost of this type of study, which requires many configurations to be assessed on many instances. To provide an idea about the cost difference, as of April 02, 2026, and according to the SWE-Bench leaderboard, the top-performing model (Claude 4.5 Opus (high reasoning)) provides 20\% more P@1 but costs 15 times more than GPT-5-mini.  

\subsubsection{Prompt Structure}

Figure~\ref{fig:prompt-template} shows the prompt template structure. The system prompt establishes the task and output format, while the user prompt template shows placeholders for the varying context sections.

\begin{figure}[t]
\begin{lstlisting}
SYSTEM PROMPT:
You are an expert software engineer.
You will receive a bug's problem statement and two tiers of context:
1) Buggy (primary) files, elements, and line ranges.
   These are confirmed edit locations. Prioritize edits here.
2) Additional relevant (secondary) files/elements/lines.
   Consider them only if needed.
Output: SEARCH/REPLACE blocks (preferred) or unified diff.
Keep the patch minimal and surgical.

USER PROMPT:
Problem statement: {problem_statement}

==== PRIMARY CONTEXT (BUGGY) ====
Buggy files: {buggy_files_list}
Buggy elements: {buggy_elements_list}
Buggy line ranges: {buggy_lines_list}

==== SECONDARY CONTEXT (RELEVANT) ====
Relevant files: {rel_files_list}
Relevant elements: {rel_elements_list}
Relevant line ranges: {rel_lines_list}

==== FILE CONTENTS (read-only) ====
-- Buggy files --
{buggy_files_section}

-- Additional relevant files --
{rel_files_section}
\end{lstlisting}
\caption{Prompt template (abbreviated). Sections are populated or left empty based on the experimental configuration.}
\label{fig:prompt-template}
\end{figure}

To isolate the effect of context variation, all configurations share a consistent prompt template comprising a system prompt and a user prompt.
The system prompt instructs the model to behave as an expert software engineer, which receives a problem statement and two tiers of context: (1) \emph{Buggy (primary) files}, \emph{elements}, and \emph{line} ranges representing confirmed edit locations, and (2) \emph{additional relevant (secondary)} context that may also be involved. The model is instructed to prioritize edits within buggy regions and to modify secondary context only when necessary. Output format is specified as either SEARCH/REPLACE blocks (preferred) or unified diff.

The user prompt contains four components: (1) the GitHub issue description (problem statement), (2) primary buggy information, which is a list of \emph{"buggy files"}, \emph{"buggy elements"}, and \emph{"buggy line"} ranges, (3) additional relevant context which is a list of relevant \emph{files}, \emph{elements}, and \emph{line} ranges (deduplicated from buggy context), and (4) the full contents of all referenced files provided as read-only material.

For each experimental configuration, we populate or omit sections accordingly. For example, in the "buggy files, No element, No line" configuration, the \emph{buggy files} list is populated while \emph{buggy elements/lines} and all relevant sections are empty. In the "LLM files, LLM element, buggy line" configuration, relevant \emph{files} and \emph{elements} are populated from the LLM retrieval, while \emph{buggy lines} provide precise line localization.

This two-tier structure (primary buggy vs. secondary relevant) allows the model to distinguish between confirmed edit locations and additional context, enabling us to study how the model utilizes different context types.

Our prompt structure is not directly adopted from a specific APR agent, but is instead a simplified and controlled design inspired by prior LLM-based repair systems such as Agentless and SWE-agent. These systems typically combine a problem statement with repository context and enforce structured output formats (e.g., patches or diffs). We intentionally avoid using full agent-style prompts (e.g., tool use, iterative reasoning, or environment interaction) in order to isolate the effect of localization context. Instead, we design a minimal prompt that retains only the essential components required for patch generation: (1) the issue description, (2) localized context, and (3) explicit output formatting constraints.

\subsection{Independent Variables}

We vary context along three orthogonal dimensions corresponding to file-, element-, and line-level reasoning. These dimensions are chosen because they capture the primary granularities at which existing LLM-based APR systems operate. Prior approaches typically retrieve and provide context at the \emph{file}-level, the \emph{element}-level, or the \emph{line}-level. While individual systems often focus on one or a subset of these granularities, our formulation unifies them into a single experimental framework. By covering all three levels, we capture the full spectrum of localization strategies used in prior work, enabling a comprehensive and systematic analysis of their individual and combined impact on repair performance.

\subsubsection{File-Level Context}

\emph{File}-level localization determines which source files are visible to the model. 
\begin{itemize}
    \item \textbf{No files:} Only the bug description is provided, with no source code.
    
    \item \textbf{Buggy files:} \emph{"Buggy files"} are the files that are in the developer's GT patch. This allows us to evaluate whether an LLM can successfully repair a bug when given the correct locations. On average, instances have 1.25 \emph{"buggy files"} (range: 1--21), totaling approximately 12,131 tokens. Using GT \emph{files} is also a very realistic setting, as recent LLM-based retrieval methods can identify relevant files with high accuracy(more than 90\%)~\cite{sepidband2026rgfl}.
    
    \item \textbf{Relevant files (rule-based):} This includes \emph{"Buggy files"} plus files that import or are imported by buggy files, which expands the context to include inter-file dependencies such as function calls, shared data structures, or incorrect API usage. We adopt an import-based expansion strategy as a lightweight and language-agnostic approximation of dependency relationships, under the assumption that imported files are more likely to contain semantically relevant definitions or usage contexts. When the entire expanded file context does not fit in the context window of GPT-5-mini (400,000), a second LLM with a larger context window (grok-4-fast-reasoning, with 2,000,000 token context size) ranks and filters less relevant files. On average, the \emph{"rule-based"} strategy results in 18.12 files (range: 0--579) per sample, approximately 96,237 tokens.
    
    \item \textbf{Relevant files (LLM):} \emph{"Buggy files"} plus relevant files in the repository that are identified by the LLM by giving the bug description. We include this strategy to capture semantic relevance that may not be reflected in explicit dependency structures such as imports. While \emph{"rule-based"} expansion approximates structural relationships, it may introduce many unrelated files or miss context that is conceptually relevant but not directly connected in the dependency graph. This yields, on average, 8.54 files (range: 1--18), approximately 58,273 tokens.
\end{itemize}

\subsubsection{Element-Level Context}

Unlike \emph{file}-level context, \emph{element}-level context depends on which files are provided to the model. Both the \emph{"Call Graph"} and \emph{"LLM element retrieval"} operate on the set of files given, yielding different element counts and token sizes for different \emph{file}-level context.

\begin{itemize}
    \item \textbf{No elements:} No element-level localization is performed.
    
    \item \textbf{Buggy elements:} only the buggy functions, classes, and global statements identified from the developer's patch. On average, 3.26 elements (range: 0--64), approximately 5,784 tokens. Unlike \emph{file}-level localization, accurately identifying such precise \emph{elements} remains more challenging for existing approaches; therefore, this setting should be interpreted as an optimistic upper bound on achievable performance.
    
    \item \textbf{Relevant elements (Call Graph):} \emph{"Buggy elements"} plus elements extracted using call graph analysis by giving the files. We construct a call graph by identifying function and method definitions and their call relationships within the provided file set. Starting from the \emph{"buggy elements"}, we traverse the call graph to collect structurally related elements, including (i) functions that are directly called by the \emph{"buggy elements"} (callees) and (ii) functions that invoke the \emph{"buggy elements"} (callers). Statistics vary by the \emph{file}-level context:
    \begin{itemize}
        \item With \emph{"buggy files"}: avg 6.5 elements (max 249), avg 1,251 tokens (max 15,143)
        \item With \emph{"rule-based files"}: avg 19.33 elements (max 1,391), avg 10,221 tokens (max 737,003)
        \item With \emph{"LLM files"}: avg 3.04 elements (max 154), avg 2,074 tokens (max 80,909)
    \end{itemize}
    
    \item \textbf{Relevant elements (LLM):} \emph{"Buggy elements"} plus elements that are identified by LLM from the provided files. We include this strategy to capture semantic relationships between code elements that may not be reflected in structural dependencies, such as the call graph. While \emph{"Call Graph"} expansion identifies execution-related connections (e.g., callers and callees), it may miss elements that are conceptually relevant to the bug, such as functions implementing similar logic, handling related edge cases, or contributing to the same higher-level functionality without direct invocation links. Statistics vary by the \emph{file}-level context:
    \begin{itemize}
        \item With \emph{"buggy files"}: avg 6.37 elements (max 27), avg 4,834 tokens (max 94,807)
        \item With \emph{"rule-based files"}: avg 20.31 elements (max 211), avg 15,110 tokens (max 127,798)
        \item With \emph{"LLM files"}: avg 18.31 elements (max 45), avg 16,375 tokens (max 167,520)
    \end{itemize}
\end{itemize}

\subsubsection{Line-Level Context}

\begin{itemize}
    \item \textbf{No lines:} No specific lines are marked as buggy.
    
    \item \textbf{Buggy lines:} \emph{"Buggy lines"} from the developer's patch. On average, 13.56 lines (range: 1--232). Precisely identifying such fine-grained locations is particularly challenging in practice; therefore, this setting should be interpreted as a strong optimistic upper bound on achievable performance.
    
    \item \textbf{Relevant lines (context window):} \emph{"Buggy lines"} plus up to $\pm$10 lines around each \emph{"buggy line"}, in the \emph{"buggy file"}. On average, 58.47 lines (range: 21--922).
    
    \item \textbf{Relevant lines (code slicing):} \emph{"Buggy lines"} plus additional lines extracted using static code slicing over the \emph{"buggy files"}. We perform slicing at the file level, restricting the analysis to the \emph{"buggy file"} to maintain consistency and avoid excessive context expansion. Starting from the buggy line ranges, we construct a lightweight dependency graph using static analysis of the program’s abstract syntax tree (AST). This graph captures (i) data dependencies via def-use relationships between variables, and (ii) control dependencies induced by constructs such as conditionals, loops, and exception handling blocks. We then compute both \emph{backward slices} (lines that influence the \emph{"buggy lines"}) and \emph{forward slices} (lines that are influenced by the \emph{"buggy lines"}), and take the union of these sets along with the original \emph{"buggy lines"}. This combined slice approximates the broader execution context in which the bug occurs, including both causes and potential downstream effects. On average, 20.24 lines (range: 1--257).
    
    \item \textbf{Relevant lines (LLM):} \emph{"Buggy lines"} plus suspicious lines that are identified by LLM. We restrict selection to the \emph{"buggy files"} to maintain consistency across configurations and to avoid introducing excessive cross-file noise at the line level. We include this strategy to capture fine-grained semantic relevance that may not be reflected in structural heuristics such as \emph{"context window"} or \emph{"static code slicing"}. On average, 743.8 lines (range: 0--209,939), though most instances have far fewer.
\end{itemize}

For LLM-based context retrieval, we use structured prompts tailored to each granularity (\emph{file}, \emph{element}, and \emph{line}). Each prompt provides the bug problem statement along with the available context (e.g., candidate files, file contents), and instructs the model to return structured outputs in JSON format (e.g., ranked file paths, relevant elements per file, or line ranges). The prompts are designed to enforce concise outputs without additional explanation, ensuring consistency across instances. At the \emph{file} level, the model ranks candidate files based on their relevance to the bug and their relationship to the \emph{"buggy files"}. At the \emph{element} and \emph{line} levels, the model identifies likely locations of the bug within the provided files. Full prompt templates are shown in Figures~\ref{fig:llm-prompt-file}, ~\ref{fig:llm-prompt-element}, and ~\ref{fig:llm-prompt-line}

\begin{figure}[t]
\scriptsize
\begin{lstlisting}
FILE-LEVEL (LLM RETRIEVAL)
Input:
- Bug problem statement
- Ground-truth (buggy) files
- Ground-truth elements and buggy line ranges
- All Python files in the repository (paths only)
Task:
- Rank repository files by relevance based on:
  (i) consistency with the bug description
  (ii) relation to the ground-truth files, elements, and lines
  (iii) file path semantics and repository structure
Output:
- JSON object with ranked file paths (most to least relevant)
\end{lstlisting}
\caption{Summary of LLM-based prompt for \emph{file} retrieval. The prompt enforces structured JSON outputs without explanations to ensure consistency and parsability.}
\label{fig:llm-prompt-file}
\end{figure}

\begin{figure}[t]
\scriptsize
\begin{lstlisting}
ELEMENT-LEVEL (LLM RETRIEVAL)
Input:
- Bug problem statement
- Source files (path + full contents)
Task:
- For each file, select the most relevant program elements:
  (functions, classes, globals)
- Prioritize elements likely to contain or cause the bug
- Limit to top-k elements per file (ranked)
Output:
- JSON mapping: file: [elements]
\end{lstlisting}
\caption{Summary of LLM-based prompt for \emph{element} retrieval. The prompt enforces structured JSON outputs without explanations to ensure consistency and parsability.}
\label{fig:llm-prompt-element}
\end{figure}

\begin{figure}[t]
\scriptsize
\begin{lstlisting}
LINE-LEVEL (LLM RETRIEVAL)
Input:
- Bug problem statement
- Full contents of buggy files
Task:
- Identify small, precise line ranges likely involved in the bug
- Prefer minimal and focused spans
Output:
- JSON mapping: file: [[start_line, end_line], ...]
\end{lstlisting}
\caption{Summary of LLM-based prompt for \emph{line} retrieval. The prompt enforces structured JSON outputs without explanations to ensure consistency and parsability.}
\label{fig:llm-prompt-line}
\end{figure}

\begin{table*}[t]
\caption{Context statistics by localization method per sample. Element-level statistics depend on which files are provided (shown in parentheses).}
\label{tab:context-stats}
\centering
\small
\begin{tabularx}{\columnwidth}{Xrrrr}
\toprule
\textbf{Context Type} & \textbf{Avg Count} & \textbf{Max Count} & \textbf{Avg Tokens} & \textbf{Max Tokens} \\
\midrule
\multicolumn{5}{l}{\textit{File-Level}} \\
Buggy files & 1.25 & 21 & 12,131 & 211,330 \\
Rule-based & 18.12 & 579 & 96,237 & 8,041,838 \\
LLM & 8.54 & 18 & 58,273 & 257,034 \\
\midrule
\multicolumn{5}{l}{\textit{Element-Level}} \\
Buggy elements & 3.26 & 64 & 5,784 & 127,443 \\
Call graph(buggy files) & 6.50 & 249 & 1,251 & 15,143 \\
Call graph(Rule files) & 19.33 & 1,391 & 10,221 & 737,003 \\
Call graph(LLM files) & 3.04 & 154 & 2,074 & 80,909 \\
LLM (buggy files) & 6.37 & 27 & 4,834 & 94,807 \\
LLM (Rule files) & 20.31 & 211 & 15,110 & 127,798 \\
LLM (LLM files) & 18.31 & 45 & 16,375 & 167,520 \\
\midrule
\multicolumn{5}{l}{\textit{Line-Level}} \\
Buggy lines & 13.56 & 232 & 112 & 1,799 \\
Context window & 58.47 & 922 & 464 & 6,885 \\
Code slicing & 20.24 & 257 & 190 & 2,784 \\
LLM & 743.80 & 209,939 & 1,179 & 211,069 \\
\bottomrule
\end{tabularx}
\end{table*}

Table~\ref{tab:context-stats} summarizes the context statistics across methods. The statistics reveal substantial variation in context size across methods. Notably, \emph{"rule-based file"} retrieval can produce extremely large contexts (up to 8M tokens), while \emph{"LLM-based file"} retrieval produces more moderate and bounded context sizes. At the \emph{element}-level, the gap between methods becomes smaller, though \emph{"LLM-based"} retrieval is generally larger than \emph{"Call Graph"} retrieval. At the \emph{line}-level, \emph{"LLM-based"} retrieval produces dramatically larger and more variable contexts (up to 209,939 lines), far exceeding all other methods. Meanwhile, \emph{"static code slicing"} remains more constrained than \emph{"context window"} expansion, producing fewer lines on average. Overall, these results highlight that LLM-based methods tend to be more selective at coarse granularity (files), but increasingly expansive at finer granularities. Therefore, both the choice of retrieval method and the level of granularity significantly affect the amount and variability of context provided to the model.

\subsection{Experimental Setup}

We evaluate all meaningful combinations of \emph{file}-, \emph{element}-, and \emph{line}-level context. Because \emph{element} and \emph{line} context require \emph{file} context to operate on, not all combinations are valid. Therefore, in total, there are 61 configurations: one baseline without localization information, and three groups of 20 configurations corresponding to \emph{"buggy files"}, \emph{"rule-based"} expansion, and \emph{"LLM-based"} expansion. Each group explores all combinations of \emph{element} and \emph{line} strategies, enabling factorial analysis of both main effects and interactions among the three levels of context expansion strategies.

For RQ1.1, RQ1.2, RQ2.1, and RQ2.2, we assess statistical significance using paired comparisons across matched configuration blocks. Specifically, for RQ1, we compare \emph{file}-level strategies across configurations with identical \emph{element}- and \emph{line}-level settings; for RQ2, we compare \emph{element}-level strategies across configurations with identical \emph{file}- and \emph{line}-level settings; and for RQ3, we compare \emph{line}-level strategies across configurations with identical \emph{file}- and \emph{element}-level settings. The resulting paired differences are non-normal (Shapiro--Wilk $p < 0.001$), so we apply the Wilcoxon signed-rank test and interpret statistical significance at $\alpha = 0.05$.

Note that although each configuration is evaluated using a single repair attempt per instance, our factorial design evaluates 61 distinct context configurations across each benchmark instance. As a result, each instance is exposed to a wide range of independent reasoning conditions across the \emph{file}-, \emph{element}-, and \emph{line}-level dimensions. The observed trends, therefore, reflect consistent behavior across many context variations rather than outcomes from a single configuration, helping reduce the impact of randomness in the aggregate analysis.

Our primary evaluation metric is the \emph{resolution rate}, defined as the proportion of instances for which the generated patch successfully passes all tests. A repair attempt is considered successful when the generated patch applies cleanly to the repository and all validation tests included in the SWE-bench test patch pass after applying the fix.

%% file: Sections/4_Results.tex
\section{Results}
\label{sec:results}

We evaluate how different localization granularities affect repair success in LLM-based program repair. Our study includes 61 configurations that vary the \emph{file}-, \emph{element}-, and \emph{line}-level context provided to the model. The following subsections are organized by research questions and analyze the impact of each granularity. 

Table~\ref{tab:all-results} summarizes the resolution results for all configurations. Each row corresponds to one configuration, defined by a combination of \emph{file}-, \emph{element}-, and \emph{line}-level context. The \textbf{File}, \textbf{Element}, and \textbf{Line} columns indicate the type of context provided at each granularity: "Buggy" refers to ground-truth context extracted from the developer patch, "Rule" denotes \emph{"rule-based"} expansion, and "LLM" refers to \emph{"LLM-based"} context retrieval. "Ctx window" corresponds to a fixed \emph{"context window"} around the target location, and "Slicing" denotes \emph{"static code slicing"} expansion. The symbol "–" indicates that no context is provided at that level. The \textbf{Resolved} column reports the number of instances (out of 500) successfully repaired under that configuration, and \textbf{Rate} shows the corresponding percentage. Row numbers are included to facilitate reference throughout the analysis.
\input{pics_tabs/all_results}

\subsection{RQ1: What is the impact of file-level context on the APR's success?}
Before analyzing the individual sub-RQs, we first examine the overall effect of introducing \emph{file}-level context. Comparing configurations that include any file context against the no-file baseline reveals a substantial performance difference. As shown in Table~\ref{tab:all-results}, the baseline configuration (row 1: No files, No elements, No lines) resolves only 18 out of 500 instances (3.6\%). In contrast, introducing file-level context immediately increases resolution rates to approximately 56--63\%, corresponding to a 15--17$\times$ improvement. 

\subsubsection{RQ1.1: Does additional file-level context, beyond the precise buggy file location(s), help LLM-based APR?}

We hypothesize that \emph{providing more file-level context than the "buggy files" improves repair performance}. To test this hypothesis, we compare configurations that expand context beyond \emph{"buggy files"} using (i) \emph{"rule-based file"} retrieval and (ii) \emph{"LLM-based file"} retrieval against \emph{"buggy file"} context, while controlling \emph{element}- and \emph{line}-level settings. This paired design isolates the effect of \emph{file}-level expansion independently of other localization granularities.

We evaluate the hypothesis using the Wilcoxon signed-rank test over paired configurations. For each comparison, configurations with identical \emph{element} and \emph{line} settings are matched across \emph{file}-retrieval strategies in Table~\ref{tab:all-results}. For example, row~2 (buggy files, no elements, no lines) is compared with row~22 (rule-based files) and row~42 (LLM-retrieved files) under the same element and line conditions. 
Similarly, row~8 (buggy files/buggy elements/buggy lines) is paired with row~28 (rule-based files/buggy elements/buggy lines) and row~48 (LLM-based files/buggy elements/buggy lines). 

The Wilcoxon signed-rank tests show statistically significant advantages for expanded \emph{file} context. Comparing \emph{"buggy files"} vs.\ \emph{"rule-based files"} yields a significant improvement ($p=0.0334$, $r=0.4201$, medium effect), while \emph{"buggy files"} vs.\ \emph{"LLM-based files"} shows a substantially stronger effect ($p=5.1\times10^{-5}$, $r=0.8682$, large effect), indicating that LLM-based expansion consistently outperforms buggy-only context.

\begin{tcolorbox}[colback=blue!5,colframe=blue!40]
\textbf{Finding 1: Expanding file context improves repair performance.}
Providing additional files beyond the buggy set generally increases repair success compared to buggy-only localization. 
\emph{"Rule-based"} expansion outperforms \emph{"buggy files"} in 14/20 configurations and \emph{"LLM-based"} expansion in 19/20 configurations, with both improvements statistically significant (Wilcoxon signed-rank test).
\end{tcolorbox}

\subsubsection{RQ1.2: Which retrieval strategy works better for file-level FL context expansion?}

We hypothesize that \emph{LLM-based file retrieval is better than rule-based}. We compare configurations that use \emph{"rule-based"} and \emph{"LLM-retrieved files"}, while controlling the \emph{element} and \emph{line} context. We evaluate the hypothesis using the Wilcoxon signed-rank test over paired configurations in Table~\ref{tab:all-results}. For example, row~22 (rule-based files, no elements, no lines) is compared with row~42 (rule-based files) under the same element and line conditions. Similarly, row~24 (rule-based files/no elements/Context Window lines) is paired with row~44 (rule-based files/buggy elements/buggy lines) and row~48 (LLM/-/Ctx). 

The Wilcoxon results show statistically significant advantages for expanding file context with LLM. Comparing \emph{"rule-based files"} vs.\ \emph{"LLM-based files"} shows a significant improvement ($p=2.71\times10^{-4}$, $r=0.7722$, large effect), indicating that \emph{"LLM-based"} expansion consistently outperforms \emph{"rule-based"} expansion.

\begin{tcolorbox}[colback=blue!5,colframe=blue!40]
\textbf{Finding 2: LLM-based file retrieval outperforms rule-based expansion, in terms of the final repair's performance.}
Across all configurations, \emph{"LLM-retrieved files"} outperform \emph{"rule-based files"} in 17/20 cases, in terms of resolve rates of the down stream repair task. The improvement is statistically significant (Wilcoxon signed-rank test). 
The improvement is strongest when \emph{line} localization is precise (\emph{"buggy lines"}), where \emph{"LLM-based files"} achieve the highest resolutions (e.g., 312 with \emph{"buggy elements"}, 309 with \emph{"Call Graph elements"}, and 317 with \emph{"LLM elements"}.).
\end{tcolorbox}


\subsubsection{RQ1.3: What is the cost of rule-based versus LLM-based file-level context expansion?}
\input{pics_tabs/RQ1.3_histogram}
\input{pics_tabs/RQ1.3_boxplots}

Figure~\ref{fig:file-histograms} shows the distribution of retrieved file counts and file-level token budgets across all instances for both \emph{"rule-based"} and \emph{"LLM-based file"} retrieval methods. As illustrated in the histograms, \emph{"LLM-based"} retrieval is concentrated in lower cost regions, with most instances clustered around small numbers of files and moderate token budgets. In contrast, \emph{"rule-based"} expansion exhibits a broader distribution, with a noticeable spread toward larger file counts and higher token budgets. This indicates that \emph{"rule-based"} methods more frequently produce large context sizes, whereas \emph{"LLM-based"} retrieval remains more tightly focused. 

This trend is further supported by the box plots in Figure~\ref{fig:file-boxplots}, which summarize the central tendency and variability of the cost distributions. \emph{"LLM-based"} retrieval shows both lower medians and narrower interquartile ranges, indicating a more compact and stable cost profile. In particular, the median number of retrieved files is 9 for \emph{"LLM-based"} methods compared to 12 for \emph{"rule-based"} methods, and the average is 8.37 versus 13.71, respectively. A similar pattern is observed for token budgets, where \emph{"LLM-based"} retrieval has a median of 41,649 tokens and an average of 50,039.80, compared to 54,360 and 62,004.83 for \emph{"rule-based"} expansion.

To improve readability and prevent extreme values from dominating the scale, statistical outliers are excluded from the plots. We identify outliers using the interquartile range (IQR) method, where $Q_1$ and $Q_3$ denote the first (25th percentile) and third (75th percentile) quartiles, respectively, and $\text{IQR} = Q_3 - Q_1$. Values falling below $Q_1 - 1.5 \times \text{IQR}$ or above $Q_3 + 1.5 \times \text{IQR}$ are removed. Since we perform a paired comparison between \emph{"rule-based"} and \emph{"LLM-based"} costs, filtering is applied jointly, retaining only instances that fall within the non-outlier range for both methods.

To evaluate the statistical significance of these differences, we perform paired Wilcoxon signed-rank tests. The results show that \emph{"LLM-based"} expansion consistently incurs lower cost than \emph{"rule-based"} expansion, with highly significant differences for both file counts ($p = 6.91 \times 10^{-21}$) and token budgets ($p = 4.65 \times 10^{-7}$). These results confirm that the observed differences are statistically robust.

The above comparison focuses on the cost of the repair phase, where the selected file context is provided to the LLM for patch generation. However, LLM-based file ranking itself incurs an additional cost, as it requires an LLM call to rank repository files, whereas rule-based ranking is almost free. Despite this overhead, the cost of LLM-based ranking is relatively small: across all 500 instances, it requires an average of 549 input tokens (median: 402), which is negligible compared to the token budgets used during repair (typically tens of thousands of tokens). Moreover, ranking is performed only once per instance and reused across all configurations, while repair is executed separately for each configuration. Since ranking is done only once, its cost is shared, and the reduction in repair cost is larger than the added ranking cost. Overall, when considering both ranking and repair phases, LLM-based file retrieval results in a lower total cost compared to rule-based expansion.

\begin{tcolorbox}[colback=blue!5,colframe=blue!40]
\textbf{Finding 3: LLM-based file retrieval is less costly compared to the rule-based expansion method, in both file and token level.}
Across all instances of SWE-bench, \emph{"LLM-based"} retrieval results in smaller context expansion, while \emph{"rule-based"} expansion's cost is higher in average and variance. The differences are statistically significant too (Wilcoxon signed-rank test). 
\end{tcolorbox}

\subsection{RQ2: What is the impact of element-level context on the APR's success?}
Before analyzing the individual sub-RQs, we first examine the overall effect of introducing \emph{element}-level context while controlling \emph{file} and \emph{line} settings. This comparison isolates the contribution of \emph{element}-level localization across different \emph{file/line} combinations in Table~\ref{tab:all-results}.




\emph{Element}-level context generally improves repair performance compared to using no \emph{elements}, but its effectiveness depends strongly on the quality of the \emph{file}-level context. When \emph{"buggy files"} are provided, adding \emph{element}-level information yields consistent gains, improving resolution by up to +20 instances across \emph{line} configurations. With \emph{"rule-based files"}, the impact becomes more variable: \emph{element} context sometimes yields substantial improvements (e.g., +21 instances with slicing and LLM-retrieved elements) but can also reduce performance due to interference from noisier \emph{file} context. In contrast, when \emph{"LLM-retrieved files"} are used, \emph{element}-level context produces the most consistent improvements, with \emph{"LLM-retrieved elements"} achieving the largest gains and enabling the best overall configuration (317 resolved instances with \emph{"buggy lines"}). Overall, these results indicate that \emph{element}-level localization provides measurable but conditional benefits, acting primarily as a complementary signal whose value depends on the precision of higher-level \emph{file} context.

\subsubsection{RQ2.1: Does additional element-level context, beyond the precise buggy element location(s), help LLM-based APR?}

We hypothesize that \emph{providing more element context than "buggy elements" helps repair}: specifically, \emph{"Call Graph elements"} and \emph{"LLM-retrieved elements"} should resolve at least as many instances as \emph{"buggy elements"} when \emph{file} and \emph{line} context are held constant. To evaluate this hypothesis, we use the Wilcoxon signed-rank test over paired configurations. We compare (i) \emph{"buggy element"} vs. \emph{"Call Graph element"} and (ii) \emph{"buggy element"} vs. \emph{"LLM element"} while controlling \emph{file} and \emph{line} context. For each comparison, configurations with identical \emph{file} and \emph{line} settings are matched across element-retrieval strategies in Table~\ref{tab:all-results}. For example, row~7 (buggy files, buggy elements, no lines) is compared with row~12 (Call Graph elements) and row~17 (LLM-retrieved elements) under the same \emph{file} and \emph{line} conditions. 

The global comparison between \emph{"Buggy element"} and \emph{"Call Graph element"} yields a Wilcoxon $p$-value of $0.9169$ and an effect size of $0.3691$ (medium), indicating no statistical difference and suggesting that \emph{"Call Graph"} expansion does not systematically improve repair performance. In fact, \emph{"Buggy elements"} outperform \emph{"Call Graph elements"} in most configurations, implying that structurally retrieved elements often introduce irrelevant context. Therefore, we also test the opposite direction, where \emph{"buggy elements"} are hypothesized to outperform \emph{"Call Graph elements"}. This test yields a $p$-value of $0.0831$, indicating a directional trend favoring \emph{"buggy elements"}, although the difference does not reach statistical significance at the $\alpha=0.05$ level.

For \emph{"Buggy elements"} vs. \emph{"LLM elements"}, the Wilcoxon test yields a $p$-value of $0.1810$ and an effect size of $0.2346$ (small). While this difference is also not statistically significant, the directional trend favors \emph{"LLM-based"} expansion, which outperforms \emph{"buggy elements"} in a majority of configurations. This suggests that semantically retrieved \emph{element} context can provide useful additional information, although the gains are modest and context-dependent.

\emph{"Call Graph elements"} outperform \emph{"buggy elements"} in only 4/15 configurations and show no overall statistical advantage. In contrast, \emph{"LLM-retrieved elements"} outperform \emph{"buggy elements"} in 10/15 configurations, indicating a consistent directional benefit despite the absence of statistical significance. Overall, these results suggest that expanding beyond \emph{"buggy elements"} can be beneficial when relevance is determined semantically (\emph{"LLM-based retrieval"}), whereas structurally retrieved expansions via call graphs frequently introduce low-signal context that limits or even reduces repair effectiveness.

The best \emph{element} context (\emph{"LLM element"}) combined with precise \emph{line} context (\emph{"buggy line"}) achieves the highest resolution across all file types: buggy files + LLM elements + buggy lines: 300 (60.0\%); Rule files + LLM elements + buggy lines: 301 (60.2\%); LLM files + LLM elements + buggy lines: \textbf{317 (63.4\%)}.
This pattern indicates that broader semantic understanding at the \emph{element}-level (via \emph{"LLM-based"} retrieval) is most effective when combined with precise, focused \emph{line}-level localization. In other words, \emph{element} expansion helps the model understand cross-function interactions and higher-level semantics, but repair performance peaks when the final edit location is sharply constrained. When \emph{line} context is imprecise (e.g., \emph{"code slicing"} or \emph{"context windows"}), the advantage of \emph{"LLM-retrieved elements"} diminishes. This highlights a complementarity principle: broad semantic context at higher levels works best when paired with precise localization at lower levels.

\begin{tcolorbox}[colback=blue!5,colframe=blue!40]
\textbf{Finding 4: LLM-based element retrieval often helps over buggy elements, whereas Call Graph retrieval generally does not.}
\emph{"LLM-based element"} expansion outperforms \emph{"buggy elements"} in 10/15 configurations, whereas \emph{"Call Graph"} expansion improves performance in only 4/15 and provides no overall advantage. The strongest results occur when \emph{"LLM-retrieved elements"} are combined with precise \emph{"buggy line"} localization, achieving the best overall configuration (317 resolved instances).
\end{tcolorbox}

\subsubsection{RQ2.2: Which retrieval strategy works better for element-level FL context expansion?}

We hypothesize that \emph{LLM-based element retrieval is more effective than rule-based expansion}: specifically, \emph{"LLM-retrieved elements"} should outperform \emph{"Call Graph elements"} across different \emph{file} and \emph{line} settings. To evaluate this hypothesis, we compare the \emph{"Call Graph"} and \emph{"LLM element"} expansion while holding the \emph{file} and \emph{line} context fixed by using the Wilcoxon signed-rank test over paired configurations in Table~\ref{tab:all-results}. For example, row~13 (buggy files, Call Graph elements, buggy lines) is compared with row~18 (buggy files, LLM elements, buggy lines).

The Wilcoxon results show a $p$-value of $0.0366$ and an effect size of $0.4782$ (medium), indicating a statistically significant advantage for \emph{"LLM-based element"} retrieval, supporting the hypothesis that semantic retrieval outperforms structurally derived expansion.

\begin{tcolorbox}[colback=blue!5,colframe=blue!40]
\textbf{Finding 5: LLM-based element expansion is more effective than Call Graph element expansion.}
Across the 15 file$\times$line configurations, \emph{"LLM-retrieved elements"} outperform \emph{"Call Graph"} expansion in 12 cases, with gains being especially consistent when combined with \emph{"LLM-retrieved files"}. The overall difference is statistically significant, and the strong directional consistency indicates that semantically guided \emph{element} retrieval provides higher-quality context than purely structural dependencies.
\end{tcolorbox}

Unlike \emph{file}-level context, \emph{element}-level expansion does not substantially increase the amount of code provided to the model. Instead of introducing additional \emph{file} contents, \emph{element}-level strategies primarily guide the model to focus on specific functions, classes, or program elements that are already present within the retrieved \emph{files}. As a result, \emph{element}-level context is incorporated as lightweight structural or semantic cues in the prompt rather than as additional code tokens. Therefore, the overall token budget remains largely unchanged across different \emph{element}-level configurations, and comparing costs at this level is not meaningful.

\subsection{RQ3: What is the impact of line-Level context on the APR's success?}
Before analyzing the individual sub-RQs, we first examine the overall effect of introducing \emph{line}-level context while keeping \emph{file} and \emph{element} settings fixed. This comparison isolates the contribution of \emph{line}-level localization across all file$\times$element combinations shown in Table~\ref{tab:all-results}.

In contrast to \emph{file}- and \emph{element}-level context, the impact of \emph{line}-level context is highly variable and frequently negative. Table~\ref{tab:line-impact} summarizes the change in resolution relative to the no-line baseline for each file$\times$element configuration. Across all combinations, \emph{"buggy lines"} produce only marginal effects on average (avg.\ $-0.8$, range $-12$ to +9), while broader expansion methods often degrade performance: \emph{"context window lines"} show an average decline of $-3.1$ (range $-21$ to +8), and \emph{"code slicing"} performs similarly poorly (avg.\ $-3.5$, range $-19$ to +7). In contrast, \emph{"LLM-retrieved lines"} yield small but more stable gains overall (avg.\ +1.1, range $-12$ to +10). Negative impacts occur in the majority of configurations for both \emph{"code slicing"} (8/12) and \emph{"context window"} expansion (7/12), with the largest degradation observed for buggy files + LLM elements + code slicing lines ($-19$) and Rule files + No elements + context window lines($-21$).

These results indicate that \emph{line}-level context behaves fundamentally differently from higher-level context. While additional \emph{files} or \emph{elements} generally provide useful semantic information, expanding \emph{line}-level context frequently introduces noise that interferes with precise localization. Consequently, \emph{line} context appears highly sensitive to quality and precision, suggesting that broader \emph{line} expansion does not reliably translate into better repair performance.
\input{pics_tabs/RQ3_compare_to_noline}

\subsubsection{RQ3.1: Does additional line-level context, beyond the precise buggy line location(s), help LLM-based APR?}

We hypothesize that \emph{providing additional line-level context beyond "buggy lines" can improve repair performance}. 
To evaluate this hypothesis, we compare \emph{"buggy line"} localization against three expansion strategies: \emph{"context windows"}, \emph{"static code slicing"}, and \emph{"LLM-retrieved lines"}, while holding \emph{file}- and \emph{element}-level context fixed. 
For each fixed \emph{(file, element)} configuration, we perform paired comparisons using the Wilcoxon signed-rank test, matching configurations that differ only in \emph{line}-retrieval strategy in Table~\ref{tab:all-results}. For example, row~8 (buggy files, buggy elements, buggy lines) is paired with row~9 (context window), row~10 (code slicing), and row~11 (LLM-retrieved lines), which share identical file and element settings.

Overall statistical results show no evidence that expanded \emph{line} context improves repair performance. The Wilcoxon test yields $p=0.8721$ and $r=0.3284$ (medium) for \emph{"buggy lines"} vs.\ \emph{"context window"}, $p=0.8452$ and $r=0.2944$ (small) for \emph{"buggy lines"} vs.\ \emph{"code slicing"}, and $p=0.4314$ and $r=0.0566$ (small) for \emph{"buggy lines"} vs.\ \emph{"LLM-retrieved lines"}, all far above the $\alpha=0.05$ significance threshold. Moreover, \emph{"buggy lines"} outperform expanded strategies in the majority of configurations. Therefore, we also test the opposite direction, where \emph{"buggy lines"} are hypothesized to outperform all the expansion strategies. The Wilcoxon test yields $p=0.1377$ for \emph{"buggy lines"} vs.\ \emph{"context window"}, $p=0.1647$ for \emph{"buggy lines"} vs.\ \emph{"code slicing"}, and $p=0.5817$ for \emph{"buggy lines"} vs.\ \emph{"LLM-retrieved lines"}, indicating the difference does not reach statistical significance at the $\alpha=0.05$ level.

\begin{tcolorbox}[colback=blue!5,colframe=blue!40]
\textbf{Finding 6: Expanded line context often harms repair performance.}
\emph{"Context window lines"} outperform \emph{"buggy lines"} in only 4/12 configurations, \emph{"code slicing"} in 4/12 configurations, and \emph{"LLM-retrieved lines"} in 5/12 configurations. None of these improvements reaches statistical significance, and \emph{"buggy lines"} remain superior in the majority of settings. Notably, \emph{"code slicing"} performs relatively better when combined with \emph{"rule-based files"}, but this effect is inconsistent across configurations. Overall, expanding beyond precise \emph{"buggy line"} localization does not reliably improve repair and frequently reduces performance.
\end{tcolorbox}

\subsubsection{RQ3.2: Which retrieval strategy works better for line-level FL context expansion?}

We hypothesize that different \emph{line}-expansion strategies vary in their effectiveness for repair: (i) \emph{"static code slicing"} should outperform fixed \emph{"context windows"} because it provides more targeted localization, and (ii) \emph{"LLM-retrieved lines"} should outperform \emph{"code slicing"} by leveraging semantic relevance. 
To evaluate these hypotheses, we directly compare the three expansion strategies \emph{("context window", "code slicing", and "LLM-retrieved lines")} while holding \emph{file}- and \emph{element}-level context fixed in Table~\ref{tab:all-results}. Configurations are paired such that only the \emph{line}-retrieval strategy differs. For example, row~14 (buggy files, Call Graph elements, context window lines) is compared with row~15 (code slicing lines) and row~16 (LLM-retrieved lines) under identical \emph{file} and \emph{element} settings.

Across configurations, the aggregate Wilcoxon signed-rank tests show that \emph{"Context Window"} and \emph{"Code Slicing"} do not differ significantly ($p = 0.6697$, $r=0.1246$, small effect), indicating that structural targeting alone does not reliably improve repair outcomes. In contrast, \emph{"LLM-retrieved lines"} significantly outperform both \emph{"Context Window"} ($p = 0.0034$, $r=0.7473$, large effect) and \emph{"Code Slicing"} ($p = 0.0254$, $r=0.5661$, large effect), suggesting that semantic relevance is more important than purely structural expansion.

\begin{tcolorbox}[colback=blue!5,colframe=blue!40]
\textbf{Finding 7: LLM-based line retrieval is a more effective expansion strategy compared to Context window and Code slicing expansions.}
\emph{"Code Slicing"} outperforms \emph{"Context Window"} in 7/12 configurations, but this advantage is not statistically significant, suggesting that purely structural expansion offers limited benefit. In contrast, \emph{"LLM-retrieved lines"} outperform \emph{"Context Window"} in 10/12 configurations and outperform \emph{"Code Slicing"} in 8/12 configurations, with both comparisons reaching statistical significance. These results indicate that semantic retrieval of \emph{line} context is more effective than either proximity-based or dependency-based expansion. Therefore, when expanding beyond \emph{"buggy lines"}, \emph{"LLM-based line" retrieval provides the most reliable gains}.
\end{tcolorbox}

Similar to the \emph{element}-level, the \emph{line}-level context does not significantly affect the overall cost of the prompt. \emph{Line}-level strategies typically provide localized guidance, such as highlighting relevant line ranges or suggesting regions of interest within already retrieved files, rather than adding substantial new code content. Consequently, the additional token overhead introduced by \emph{line}-level context is minimal compared to \emph{file}-level expansion. For this reason, we do not perform a cost analysis at the \emph{line}-level and instead focus on its impact on repair performance.

\noindent\textbf{Why does broader File/Element context help (but not Lines)?}
We speculate that broader \emph{file/element} context helps the model understand how buggy code fits into the system, coding conventions, and similar code that might serve as repair templates. Many bugs require understanding callers/callees that buggy localization misses. At the \emph{line}-level, the repair task shifts from "understand" to "edit". Here, additional lines introduce noise (irrelevant code paths, error handling, or logging) that dilutes the signal. The model already has \emph{file}-level context for understanding; line context should focus attention on edit locations, not expand it.

\subsection{RQ4: What are the characteristics of the best- and worst-performing FL context expansion strategies?}

We conducted a cross-dimensional analysis to examine how \emph{file}-, \emph{element}-, and \emph{line}-level contexts interact and which combinations yield the strongest and weakest overall repair performance. Several patterns emerge across configurations. To understand these patterns, we analyze the characteristics shared by the best-performing and worst-performing setups and how different context types interact across the three localization levels.

Table~\ref{tab:all-results} lists all configurations by resolution rate. The top-performing configurations are:

\begin{enumerate}
    \item Row 58: LLM files + LLM elements + Buggy lines --- 317 (63.4\%)
    \item Row 48: LLM files + Buggy elements + Buggy lines --- 312 (62.4\%)
    \item Row 49: LLM files + Buggy elements + Context window lines --- 311 (62.2\%)
    \item Row 61: LLM files + LLM elements + LLM lines --- 310 (62.0\%)
    \item Row 51: LLM files + Buggy elements + LLM lines --- 309 (61.8\%)
    \item Row 53: LLM files + Call Graph elements + Buggy lines --- 309 (61.8\%)
    \item Row 57: LLM files + LLM elements + No lines --- 308 (61.6\%)
    \item Row 60: LLM files + LLM elements + Code slicing lines --- 307 (61.4\%)
    \item Row 56: LLM files + Call Graph elements + LLM lines --- 306 (61.2\%)
    \item Row 52: LLM files + Call Graph elements + No lines --- 304 (60.8\%)
\end{enumerate}

In contrast, the lowest-performing configurations (excluding the baseline) are: 
\begin{enumerate}
    \item Row 24: Rule-based files + No elements + Context window lines --- 276 (55.2\%)
    \item Row 5: Buggy files + No elements + Code slicing lines --- 280 (56\%)
    \item Row 2: Buggy files + No elements + No lines --- 281 (56.2\%)
    \item Row 3: Buggy files + No elements + Buggy lines --- 282 (56.4\%)
    \item Row 20: Buggy files + LLM elements + Code slicing lines --- 282 (56.4\%)
    \item Row 25: Rule-based files + No elements + Code slicing lines --- 283 (56.6\%)
    \item Row 10: Buggy files + Buggy elements + Code slicing lines --- 284 (56.8\%)
    \item Row 28: Rule-based files + Buggy elements + Buggy lines --- 284 (56.8\%)
    \item Row 13: Buggy files + Call Graph elements + Buggy lines --- 285 (57\%)
    \item Row 11: Rule-based files + No elements + LLM lines --- 288 (57.6\%)
\end{enumerate}

Several consistent patterns emerge from this ranking:

\noindent\textbf{LLM-retrieved files dominate top-performing configurations.}
All top 10 configurations rely on \emph{"LLM-based file"} retrieval, highlighting the importance of semantically guided repository context. The first configuration without \emph{"LLM-retrieved files"} appears only at rank 13 (\emph{"rule-based files"} with \emph{"LLM elements"} and \emph{"code slicing lines"}, 304/500), indicating a clear advantage for \emph{"LLM-based file"} expansion.

\noindent\textbf{Absence of element context consistently harms performance.}
Many of the lowest-performing configurations omit \emph{element}-level localization. Without function or structural guidance, the model lacks information about relationships between code components, making it more difficult to reason about dependencies and generate correct edits.

\noindent\textbf{Buggy file localization is often too narrow.}
Most of the worst combinations use \emph{"Buggy files"}. \emph{"Buggy files"} alone (precise but narrow) may miss related files needed for repair, and are too focused and lack broader context. \emph{"Buggy files"} need complementary context \emph{(elements/lines)} to work well.

\noindent\textbf{Code slicing frequently appears in the worst-performing configurations.}
Four of the ten worst-performing configurations include \emph{"Code Slicing lines"}. Although slicing is designed to produce precise \emph{line}-level signals, it can conflict with the surrounding context provided at higher levels of localization. In particular, slicing-derived lines may introduce a precision mismatch when the \emph{file}-level context is limited (e.g., \emph{"buggy files"} only), restricting the model’s ability to reason about broader dependencies. Our results suggest that slicing performs better when paired with expanded file context, such as \emph{"Rule-based files"}.

\noindent\textbf{Hybrid strategies outperform uniform localization.}
The best-performing configuration (row~58, 317/500) combines \emph{"LLM-retrieved files"}, \emph{"LLM-retrieved elements"}, and precise \emph{"Buggy line"} localization. In comparison, fully precise localization (buggy/buggy/buggy) resolves 299 instances, while fully \emph{"LLM-based"} localization resolves 310. These results suggest that combining broad semantic context at higher granularities with precise \emph{line}-level localization yields the strongest repair performance.

\noindent\textbf{Buggy line combined with LLM-retrieved files consistently yields the strongest performance across all element configurations.}
When \emph{line}-level localization is precise (\emph{"buggy lines"}), configurations using \emph{"LLM-retrieved files"} achieve the highest resolution regardless of the \emph{element} strategy. Specifically, LLM files + buggy elements + buggy lines resolves 312 instances, LLM files + Call Graph elements + buggy lines resolves 309 instances, and LLM files + LLM elements + buggy lines achieves the overall best result of 317 resolved instances. This pattern indicates that broad, semantically retrieved \emph{file} context works best when paired with precise \emph{line} localization, while the choice of \emph{element} expansion plays a secondary role.

\noindent\textbf{The complementarity principle: The most effective pattern combines broad file/element context with precise line localization.}
When the \emph{file} and \emph{element} contexts are already broad, adding more \emph{line} context \emph{("context window", "code slicing", "LLM lines")} provides little benefit or even harms. The model benefits from a broad context for understanding but precise localization for focus.

%% file: pics_tabs/all_results.tex
\begin{table*}[t]
\renewcommand{\arraystretch}{0.8}
\setlength{\aboverulesep}{0pt}
\setlength{\belowrulesep}{0pt}
\setlength{\tabcolsep}{3pt}
\small
\caption{Resolution rates for all 61 experimental configurations. Rows are numbered for reference.}
\label{tab:all-results}
\centering
\begin{tabular}{clllrr}
\toprule
\textbf{\#} & \textbf{File} & \textbf{Element} & \textbf{Line} & \textbf{Resolved} & \textbf{Rate} \\
\midrule
1  & -    & -   & -           & 18  & 3.6\% \\
\midrule
2  & Buggy   & -   & -           & 281 & 56.2\% \\
3  & Buggy   & -   & Buggy          & 282 & 56.4\% \\
4  & Buggy   & -   & Ctx window  & 289 & 57.8\% \\
5  & Buggy   & -   & Slicing     & 280 & 56.0\% \\
6  & Buggy   & -   & LLM         & 291 & 58.2\% \\
7  & Buggy   & Buggy  & -           & 295 & 59.0\% \\
8  & Buggy   & Buggy  & Buggy          & 299 & 59.8\% \\
9  & Buggy   & Buggy  & Ctx window  & 297 & 59.4\% \\
10 & Buggy   & Buggy  & Slicing     & 284 & 56.8\% \\
11 & Buggy   & Buggy  & LLM         & 301 & 60.2\% \\
12 & Buggy   & Call Graph & -        & 292 & 58.4\% \\
13 & Buggy   & Call Graph & Buggy       & 285 & 57.0\% \\
14 & Buggy   & Call Graph & Ctx window & 290 & 58.0\% \\
15 & Buggy   & Call Graph & Slicing  & 296 & 59.2\% \\
16 & Buggy   & Call Graph & LLM      & 295 & 59.0\% \\
17 & Buggy   & LLM & -            & 301 & 60.2\% \\
18 & Buggy   & LLM & Buggy           & 300 & 60.0\% \\
19 & Buggy   & LLM & Ctx window   & 292 & 58.4\% \\
20 & Buggy   & LLM & Slicing      & 282 & 56.4\% \\
21 & Buggy   & LLM & LLM          & 289 & 57.8\% \\
\midrule
22 & Rule & -   & -            & 297 & 59.4\% \\
23 & Rule & -   & Buggy           & 295 & 59.0\% \\
24 & Rule & -   & Ctx window   & 276 & 55.2\% \\
25 & Rule & -   & Slicing      & 283 & 56.6\% \\
26 & Rule & -   & LLM          & 288 & 57.6\% \\
27 & Rule & Buggy  & -            & 296 & 59.2\% \\
28 & Rule & Buggy  & Buggy           & 284 & 56.8\% \\
29 & Rule & Buggy  & Ctx window   & 296 & 59.2\% \\
30 & Rule & Buggy  & Slicing      & 302 & 60.4\% \\
31 & Rule & Buggy  & LLM          & 301 & 60.2\% \\
32 & Rule & Call Graph & -       & 299 & 59.8\% \\
33 & Rule & Call Graph & Buggy      & 301 & 60.2\% \\
34 & Rule & Call Graph & Ctx window & 288 & 57.6\% \\
35 & Rule & Call Graph & Slicing & 303 & 60.6\% \\
36 & Rule & Call Graph & LLM     & 297 & 59.4\% \\
37 & Rule & LLM & -            & 297 & 59.4\% \\
38 & Rule & LLM & Buggy           & 301 & 60.2\% \\
39 & Rule & LLM & Ctx window   & 298 & 59.6\% \\
40 & Rule & LLM & Slicing      & 304 & 60.8\% \\
41 & Rule & LLM & LLM          & 300 & 60.0\% \\
\midrule
42 & LLM  & -   & -            & 303 & 60.6\% \\
43 & LLM  & -   & Buggy           & 292 & 58.4\% \\
44 & LLM  & -   & Ctx window   & 297 & 59.4\% \\
45 & LLM  & -   & Slicing      & 298 & 59.6\% \\
46 & LLM  & -   & LLM          & 303 & 60.6\% \\
47 & LLM  & Buggy  & -            & 304 & 60.8\% \\
48 & LLM  & Buggy  & Buggy           & 312 & 62.4\% \\
49 & LLM  & Buggy  & Ctx window   & 311 & 62.2\% \\
50 & LLM  & Buggy  & Slicing      & 299 & 59.8\% \\
51 & LLM  & Buggy  & LLM          & 309 & 61.8\% \\
52 & LLM  & Call Graph & -       & 304 & 60.8\% \\
53 & LLM  & Call Graph & Buggy      & 309 & 61.8\% \\
54 & LLM  & Call Graph & Ctx window & 302 & 60.4\% \\
55 & LLM  & Call Graph & Slicing & 295 & 59.0\% \\
56 & LLM  & Call Graph & LLM     & 306 & 61.2\% \\
57 & LLM  & LLM & -            & 308 & 61.6\% \\
58 & LLM  & LLM & Buggy           & \textbf{317} & \textbf{63.4\%} \\
59 & LLM  & LLM & Ctx window   & 304 & 60.8\% \\
60 & LLM  & LLM & Slicing      & 307 & 61.4\% \\
61 & LLM  & LLM & LLM          & 310 & 62.0\% \\
\bottomrule
\end{tabular}
\end{table*}

%% file: pics_tabs/RQ1.3_histogram.tex
\begin{figure*}[t]
    \centering
    \includegraphics[width=0.95\textwidth]{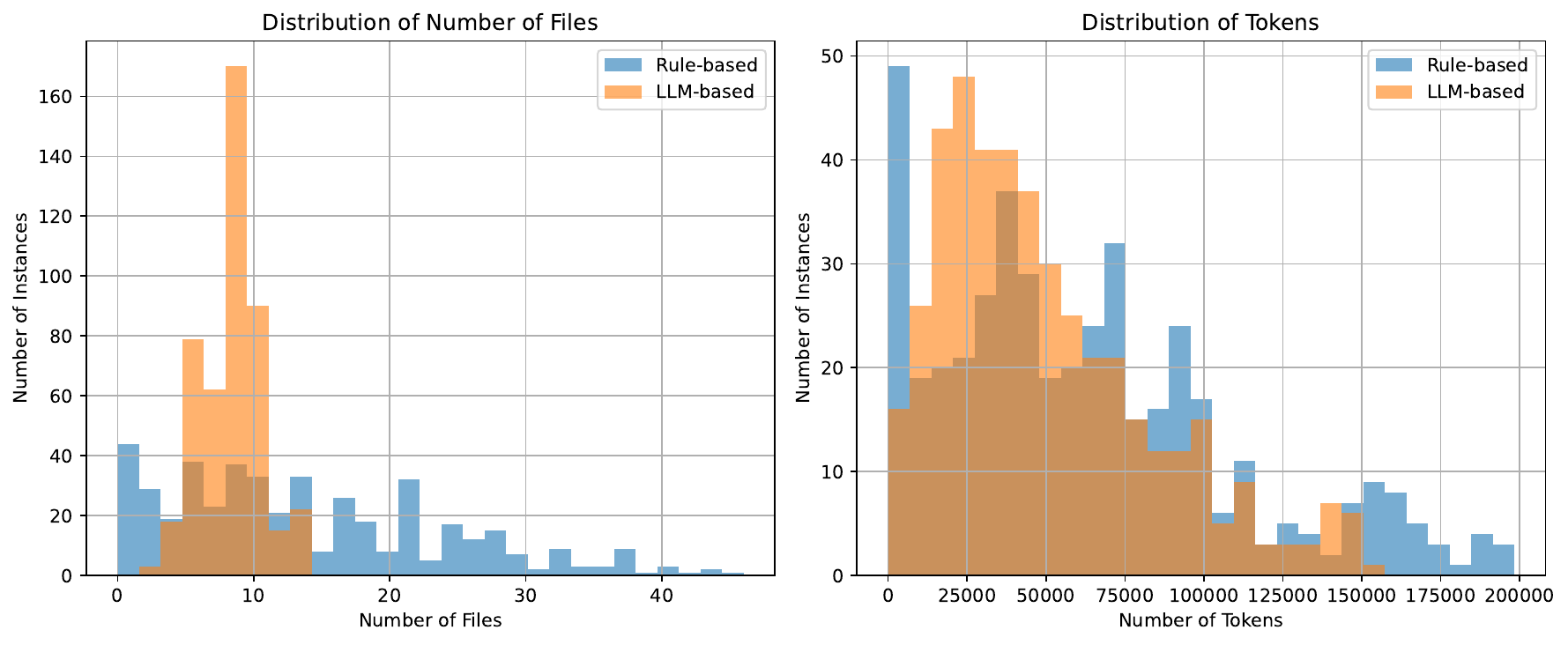}
    \caption{Distribution of file counts and file-level token budgets for all instances under different file-selection methods. 
    }
    \label{fig:file-histograms}
\end{figure*}

%% file: pics_tabs/RQ1.3_boxplots.tex
\begin{figure*}[t]
    \centering
    \includegraphics[width=0.95\textwidth]{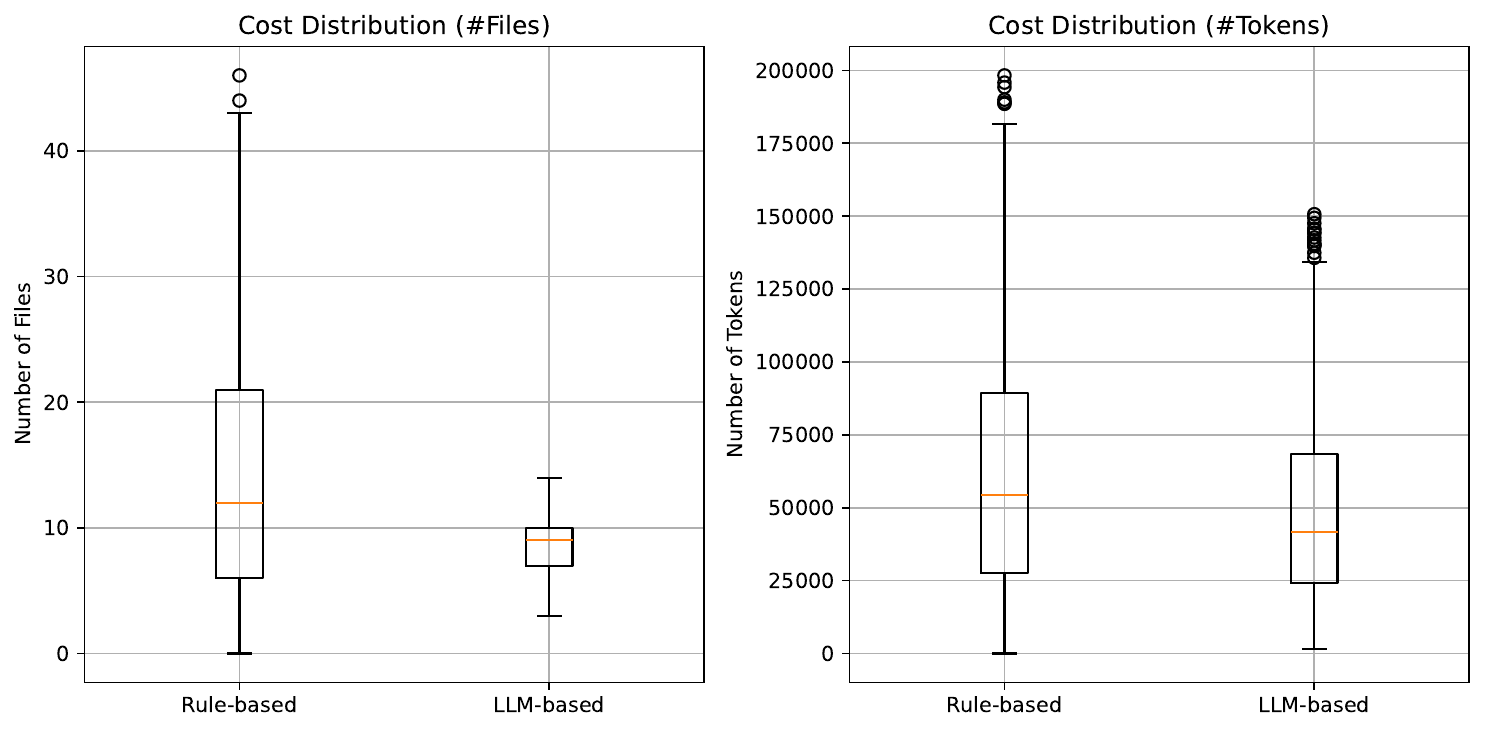}
    \caption{Box plots of file counts and file-level token budgets for all instances under different file-selection methods. 
    }
    \label{fig:file-boxplots}
\end{figure*}

%% file: pics_tabs/RQ3_compare_to_noline.tex
\begin{table}[t]
\caption{Line context impact across configurations (change vs. no line)}
\label{tab:line-impact}
\centering
\small
\begin{tabular}{llrrrr}
\toprule
\textbf{Files} & \textbf{Elements} & \textbf{GT} & \textbf{Ctx} & \textbf{Slice} & \textbf{LLM} \\
\midrule
\multirow{4}{*}{GT} 
 & No elem & +1 & +8 & $-1$ & +10 \\
 & GT elem & +4 & +2 & $-11$ & +6 \\
 & Rel (rule) & $-7$ & $-2$ & +4 & +3 \\
 & Rel (LLM) & $-1$ & $-9$ & $-19$ & $-12$ \\
\midrule
\multirow{4}{*}{Rule} 
 & No elem & $-2$ & $-21$ & $-14$ & $-9$ \\
 & GT elem & $-12$ & 0 & +6 & +5 \\
 & Rel (rule) & +2 & $-11$ & +4 & $-2$ \\
 & Rel (LLM) & +4 & +1 & +7 & +3 \\
\midrule
\multirow{4}{*}{LLM} 
 & No elem & $-11$ & $-6$ & $-5$ & 0 \\
 & GT elem & +8 & +7 & $-5$ & +5 \\
 & Rel (rule) & +5 & $-2$ & $-9$ & +2 \\
 & Rel (LLM) & +9 & $-4$ & $-1$ & +2 \\
\bottomrule
\end{tabular}
\end{table}

%% file: Sections/5_Qualitative_analysis.tex
\section{Qualitative Analysis}
\label{sec:analysis}

To better understand the mechanisms underlying our quantitative findings, we conducted a systematic qualitative analysis of instances where both more context helped and did not help across all three granularities. We examined representative cases and categorized recurring patterns based on their root causes and model behavior. These categories highlight the structural differences between configurations that improve repair performance and those that degrade it. For each category, we present a representative example to illustrate the underlying mechanism.

Across all 177 comparisons, 61 violate our RQ$i$.1 and RQ$i$.2 hypotheses, meaning that providing additional context degraded repair performance. The remaining 116 comparisons confirm our hypotheses, indicating that additional context improved or maintained repair success. To understand these outcomes, we qualitatively examine both violated and confirming cases, grouping them into recurring patterns that reveal the underlying causes of success and failure.

\paragraph{\textbf{Categories of Violated Cases (where more context did not help):}}
Violations consistently arise when additional context introduces information that is irrelevant, misleading, or unnecessarily broad for the underlying fix. In these cases, the model’s reasoning is diverted away from the true fault location or the correct repair strategy. Across \emph{file}-, \emph{element}-, and \emph{line}-level expansions, we identify four recurring categories:

\paragraph{\textbf{(V1) Excessive Context.}}
In several cases, the provided context is excessively large, overwhelming the model and shifting its attention away from the true fault location. Table~\ref{tab:v1-sympy21379} illustrates a representative violation (\path{sympy__sympy-21379}), where a substitution operation unexpectedly raises a \texttt{PolynomialError} due to an unguarded \texttt{gcd} call.
Under \emph{"LLM-based"} expansion, additional related files cause the model to modify \path{sympy/polys/polytools.py}, which appears in the expanded context but is not part of the \emph{"rule-based"} relevant set. Rather than addressing the consumer bug in \texttt{Mod.eval}, the model fixes a symptom by changing how \texttt{polytools} raises exceptions for \texttt{Piecewise} generators. The excess surrounding library code encourages over-generalization and the model reframes the problem as a global API inconsistency rather than a localized guard omission.

\input{pics_tabs/V1}

\paragraph{\textbf{(V2) Irrelevant Context Interference.}}

Additional context can introduce semantically related but non-causal code that distracts the model from the true bug. In \path{sympy__sympy-11618}, the bug arises because distance computation between points of different dimensions ignores additional coordinates when the arguments are zipped together. For example, \texttt{Point(2,0).distance(Point(1,0,2))} incorrectly returns \texttt{1} instead of $\sqrt{5}$, as the third coordinate is silently discarded.
Both the \emph{"buggy element"} and expanded-element configurations modify the same \emph{file}, \emph{element}, and \emph{lines}. However, when additional geometry-related functions(e.g., \texttt{Point2D}, \texttt{Point3D}) are included in context, the model incorrectly infers that dimensional mismatch should be treated as a strict type violation. As a result, it raises a \texttt{ValueError} when vectors differ in dimension, whereas the buggy element patch correctly promotes the lower-dimensional point by padding missing coordinates with zeros. Here, unrelated abstractions bias the model toward an overly restrictive interpretation rather than correcting the coordinate handling logic.

\paragraph{\textbf{(V3) Unnecessary Context for Localized Bugs.}}
For simple bugs requiring small localized edits, additional context may encourage unnecessary modifications. In \path{django__django-11299}, a database migration fails because Django generates incorrect SQL for a \texttt{CheckConstraint} containing mixed \texttt{AND}/\texttt{OR} clauses, incorrectly inserting fully qualified column names during table rewriting. The correct fix requires a small, localized adjustment to constraint generation rather than broader architectural changes. Under the \emph{"buggy file"} context, the model produces a minimal and correct patch. However, when additional related files are provided, the model rewrites a larger surrounding block and introduces indentation drift, breaking otherwise correct logic. The expanded context encourages unnecessary refactoring, increasing the risk of syntactic and structural errors.

\paragraph{\textbf{(V4) Over-Generalized Fixes.}}
Broader context can cause the model to propose overly ambitious architectural changes instead of localized fixes. In \path{sympy__sympy-20801}, the bug concerns inconsistent equality behavior between numeric zero and the Boolean constant \texttt{S.false}. Specifically, \texttt{S(0.0) == S.false} evaluates to \texttt{True}, while the reversed comparison yields \texttt{False}, leading to asymmetric and unintuitive results. The intended behavior is to ensure consistent equality semantics between numeric falsy values and Boolean constants. With limited context, the model produces a narrowly scoped fix that correctly handles numeric falseness in the relevant comparison path. However, with a broader context spanning Boolean logic internals, the model attempts to redefine equality semantics across Boolean classes, introducing a complex but incorrect architectural change. The broader patch reflects speculative redesign rather than targeted bug fixing.

\paragraph{\textbf{Categories of Non-Violated Cases((where more context help)):}}
Configurations that confirm our hypotheses share a key property: the additional context is \emph{necessary} to correctly understand or implement the fix. We identify four dominant categories:

\paragraph{\textbf{(N1) Cross-Component Dependencies.}}
Some bugs inherently require reasoning across multiple modules or abstractions. Table~\ref{tab:n1-django12774} shows a representative example (\path{django__django-12774}), where \path{QuerySet.in_bulk()} incorrectly rejects fields that are unique through \texttt{UniqueConstraint} metadata. Under \emph{"buggy file"} context, the model applies a shortcut heuristic by inspecting generic constraints, which fails to capture Django’s canonical uniqueness semantics. Providing \emph{"rule-based"} related files exposes the framework mechanism (\path{total_unique_constraints}), anchoring the model to the intended abstraction and producing a correct and robust fix.

\input{pics_tabs/N1}

\paragraph{\textbf{(N2) Improved Semantic Understanding.}}
In some cases, additional element context helps the model understand the intended processing pipeline and distinguish between superficial symptoms and the true semantic cause of the bug. For example, In \path{sphinx-doc__sphinx-9602}, the bug occurs because Sphinx treats literal values inside \texttt{typing.Literal} annotations (e.g., \texttt{Literal[True]}) as cross-referenceable Python classes. When nitpicky mode is enabled, this causes warnings such as ``missing \texttt{py:class}'' for values like \texttt{True}, which are not classes.
With \emph{"buggy element"} context, the model applies a superficial fix by rendering literal values as \texttt{nodes.literal}, preventing cross-referencing but altering intended formatting behavior. However, when additional related elements are provided, the model recognizes that literal handling and cross-reference generation are coordinated across annotation parsing and identifier resolution logic. The expanded context reveals that literal values should be distinguished from identifiers during cross-reference conversion rather than rendered differently at parse time. As a result, the model produces an architecture-consistent fix that preserves formatting while preventing incorrect cross-references.

\paragraph{\textbf{(N3) Context Matching Bug Complexity.}}
For genuinely complex bugs, broader context helps preserve algorithmic intent. In \path{sympy__sympy-13877}, the bug arises when determinant computation using the Bareiss algorithm produces \texttt{NaN} values for certain symbolic matrices, eventually triggering an \texttt{Invalid NaN comparison} during simplification. Correct behavior, therefore, requires understanding how symbolic cancellation interacts with recursive determinant evaluation rather than removing the simplification step altogether.
With \emph{"Buggy file"} context, the model removes the \texttt{cancel(ret)} call entirely, avoiding the crash but breaking an essential simplification step.
With \emph{"rule-based related files"}, the model instead diagnoses the real issue:
cancellation is correct but fragile in corner symbolic cases. It therefore guards the operation and correctly reuses its return value (\texttt{ret = cancel(ret)}), preserving correctness while handling the edge case.

\paragraph{\textbf{Key Insight:}}
Across all three granularities, hypothesis violations consistently arise when the provided context does not match the true semantic requirements of the bug. Broader context improves repair when cross-file dependencies or multi-function reasoning are necessary, but degrades performance when fixes are localized or when irrelevant information alters the model’s reasoning trajectory. Importantly, the violated and non-violated cases reveal complementary strengths: instances harmed by broader context are often those that benefit most from precise localization, while complex cross-module bugs require expansion to expose missing semantic signals.

These observations suggest that context selection should not be static. 
Rather than relying on a single localization configuration, APR systems may benefit from adaptive or ensemble strategies that vary context size and granularity based on characteristics of each bug rather than relying on a single fixed localization strategy. Such approaches could leverage complementary failure modes across configurations, improving overall repair robustness while avoiding unnecessary context expansion.

%% file: pics_tabs/V1.tex
\begin{figure*}
\centering
\scriptsize
\renewcommand{\arraystretch}{1.05}
\begin{tabular}{p{0.45\linewidth} p{0.45\linewidth}}
\toprule
\textbf{Rule-Based Context (Correct Patch)} &
\textbf{LLM-Based Context (Incorrect Patch)} \\
\midrule
\textbf{File Modified:}

\texttt{sympy/core/mod.py} 

\textcolor{green!60!black}{Correct consumer location}
&
\textbf{File Modified:}

\texttt{sympy/polys/polytools.py}

\textcolor{red!70!black}{Unrelated library file}
\\
\midrule
\begin{lstlisting}[language=diff,basicstyle=\ttfamily\scriptsize]

- G = gcd(p, q)
+ try:
+     G = gcd(p, q)
+ except PolynomialError:
+     # handle non polynomial cases
+     G = S.One

\end{lstlisting}
&
\begin{lstlisting}[language=diff,basicstyle=\ttfamily\scriptsize]
if isinstance(k, Piecewise):
- raise PolynomialError(
-   "Piecewise generators..."
- )
+ raise PolificationFailed(
+   opt, origs, exprs
+ )
\end{lstlisting}
\\
\midrule
\textcolor{green!60!black}{
\textbf{Localized Fix.}
Guards the \texttt{gcd} call inside
\texttt{Mod.eval}, directly addressing
the consumer failure.}
&
\textcolor{red!70!black}{
\textbf{Symptom Fix.}
Changes exception behavior inside
\texttt{polytools}, reframing the issue
as a global API problem.}
\\
\bottomrule
\end{tabular}
\caption{
Example of V1-Excessive Context (sympy\_\_sympy-21379).
The bug arises when subs() triggers a PolynomialError from gcd during the evaluation of Piecewise expressions. Rule-based file selection produces a localized fix in the correct consumer file, while LLM-based expansion introduces excessive context and causes the model to modify an unrelated library file.
}
\label{tab:v1-sympy21379}
\end{figure*}

%% file: pics_tabs/N1.tex
\begin{figure*}[t]
\centering
\scriptsize
\renewcommand{\arraystretch}{1.05}
\begin{tabular}{p{0.45\linewidth} p{0.45\linewidth}}
\toprule
\textbf{Buggy Context (Incorrect Patch)} &
\textbf{Rule-Based Context (Correct Patch)} \\
\midrule
\textbf{File Modified:}

\texttt{django/db/models/query.py}

\textcolor{red!70!black}{Heuristic uniqueness check}

&
\textbf{File Modified:}

\texttt{django/db/models/query.py}

\textcolor{green!60!black}{Framework abstraction used}
\\
\midrule
\begin{lstlisting}[language=diff,basicstyle=\ttfamily\scriptsize]

+ for constraint in
+ self.model._meta.constraints:
+     if getattr(constraint,
+        'fields',None) and
+        field_name in
+        constraint.fields:
+         is_unique=True

\end{lstlisting}
&
\begin{lstlisting}[language=diff,basicstyle=\ttfamily\scriptsize]
+ for constraint in getattr(
+     self.model._meta,
+     'total_unique_constraints',
+     ()
+ ):
+     if fields and len(fields)==1
+        and fields[0]==field_name:
+         is_unique_via_constraint=True
\end{lstlisting}
\\
\midrule
\textcolor{red!70!black}{
\textbf{Shortcut Reasoning.}
Directly scans constraints, ignoring Django’s
canonical uniqueness abstraction.}
&
\textcolor{green!60!black}{
\textbf{Cross-File Reasoning.} Uses \texttt{total\allowbreak\_unique\allowbreak\_constraints}, capturing framework-level semantics exposed through additional related files.}
\\
\bottomrule
\end{tabular}
\caption{Example of N1-Cross-Component Dependencies (django\_\_django-12774). QuerySet.in\_bulk() incorrectly rejects fields that are unique via UniqueConstraint. Buggy context produces a heuristic constraint inspection, while rule-based context exposes Django’s intended abstraction (total\_unique\_constraints), enabling a correct fix.}
\label{tab:n1-django12774}
\end{figure*}

%% file: Sections/6_Discussion.tex
\section{Practical Implications}
\label{sec:discussion}

We observed several consistent patterns across our experiments that reveal how different types of localization context influence repair success. These patterns suggested that not all context is equally useful, and that more context is not necessarily better. To better understand this behavior, we analyzed how different granularities interact, how much context is actually needed, and where additional information begins to hurt rather than help. This analysis led to a set of practical insights for designing effective context retrieval strategies in LLM-based APR systems.

\subsubsection{Prioritize File-Level Localization}

\emph{File}-level localization provides by far the largest return on investment. APR systems should (1) invest effort in accurate \emph{file}-level FL before finer-grained localization, (2) consider using \emph{"LLM-based file"} retrieval, which outperforms both \emph{"rule-based"} expansion and \emph{"buggy file"} approaches.

\subsubsection{Keep Line-Level Context Minimal}

Unlike \emph{files} and \emph{elements}, more \emph{line}-level context often degrades the performance. Methods that provide more lines (especially \emph{"context windows"} and \emph{"code slicing"}) consistently underperform precise \emph{"Buggy lines"} or even no line context at all. Additional lines introduce noise (irrelevant code paths, error handling, logging) that dilutes the repair signal. When the model already has \emph{file}-level context for understanding, \emph{line} context should be precise and focused, not expanded. When \emph{line}-level localization is available, it is better to use it precisely, but when we don't have the precise lines, providing no line context often outperforms attempting to infer relevant lines.

\subsubsection{Use Complementary Context Combinations}

The strong interaction effects observed across granularity levels suggest that fault localization context should be selected to complement rather than duplicate information across dimensions. Broad contextual expansion is most effective when paired with precise localization signals that maintain focus. For example, when \emph{file}-level context is expanded through \emph{"LLM-retrieved files"}, precise \emph{line}-level localization (e.g., \emph{"Buggy lines"}) helps constrain the model’s attention to the true fault location.

These results indicate that uniformly broad or uniformly narrow context across all dimensions is suboptimal. Instead, successful configurations balance coverage and precision across granularities. The best-performing configuration \emph{("LLM files" + "LLM elements" + "Buggy lines")} exemplifies this principle, combining broad semantic context for understanding with precise localization signals that guide repair toward the correct edit location.

\subsubsection{Prefer LLM-Based File Retrieval over Heuristic Expansion}

Our findings suggest that APR systems should prioritize \emph{"LLM-based file"} retrieval over heuristic (\emph{"rule-based"}) expansion when selecting relevant files. In RQ1.2, \emph{"LLM-based"} retrieval achieves higher repair performance compared to \emph{"rule-based"} methods. In RQ1.3, we further show that \emph{"LLM-based"} retrieval incurs lower cost, both in terms of the number of retrieved files and the associated token budgets. 

Taken together, these results indicate that \emph{"LLM-based"} retrieval provides more effective context while using less information, making it a more suitable strategy for \emph{file} selection in repository-level APR. Rather than relying on structural heuristics such as import-based expansion.

\subsubsection{Ensemble and Adaptive Strategies}

Our qualitative analysis shows that no single context configuration is consistently optimal across all bugs. Different configurations succeed on complementary subsets of instances because failures often arise from distinct mechanisms. For example, broader context improves performance when bugs require cross-file or cross-element reasoning, but harms repair when it introduces irrelevant abstractions or encourages over-generalization. Conversely, highly localized context is effective for simple or well-isolated bugs but may lack the semantic coverage required for complex fixes.

These complementary strengths suggest that APR systems should avoid static context retrieval strategies. Instead, ensemble or adaptive approaches that attempt repair under multiple context configurations may substantially improve overall resolution. Systems could dynamically adjust context granularity based on signals such as bug scope, dependency structure, or early reasoning failures, selecting a broader context for semantically complex bugs and a narrower context for localized ones. Such adaptive context provisioning allows systems to capture the benefits of expanded reasoning while minimizing distraction from irrelevant code.

%% file: Sections/7_Relatedwork.tex
\section{Related Work}
\label{sec:related_work}

\subsection{Context Engineering for Repository-Level Repair}

LLM-based repair systems operate under strict context window constraints, requiring careful selection and organization of repository artifacts. 
In repository-level issue resolution benchmarks such as SWE-bench~\cite{jimenez2023swe}, success depends not only on generation capability but also on providing the model with appropriate cross-module evidence without overwhelming it. 

This has given rise to \emph{context engineering}: the design space of (i) selecting and expanding evidence (files, functions/classes, line ranges), and (ii) packaging that evidence into prompts through ordering, prioritization, constraints, and edit formatting.

A large body of work in repository-scale systems relies on retrieval to choose a set of relevant files before generation~\cite{zhang2024autocoderover,xia2024agentless,sepidband2026rgfl}.
Common strategies include lexical retrieval (e.g., BM25-style ranking)~\cite{zhou2012should}, dense retrieval via code/text embeddings~\cite{guo2020graphcodebert}, and hybrid reranking pipelines~\cite{zhang2024autocoderover}. Systems often combine retrieval with structural heuristics such as import graphs, dependency neighborhoods, or test-failure traces to find contexts that are related to the fault region~\cite{chen2024hetfl,wang2024entity,prenner2025simple}.

Beyond files, many pipelines extract salient program elements (functions, classes, global variables) to reduce noise and focus reasoning~\cite{yu2025orcaloca}.
Structural approaches use call graphs or symbol dependencies to expand candidate elements~\cite{chen2025locagent}, while semantic approaches use LLM to nominate relevant entities based on the issue description and repository context~\cite{xia2024agentless,sepidband2026rgfl,wang2021beep}. This element-level packaging is closely related to research on code search and code representation models~\cite{guo2020graphcodebert,husain2019codesearchnet}, which aim to bridge natural language intent and program structure.

At the finest granularity, systems highlight or extract line ranges around suspicious locations using context windows~\cite{xia2024agentless}, dynamic slicing~\cite{soremekun2021locating}, or LLM-selected spans~\cite{wang2021beep}. Unlike higher-level context, line context is often used not to broaden understanding but to constrain editing and reduce unintended modifications.

\textbf{Prompt and interface designs for reliable editing.}
Work on LLM-based software agents has shown that how context is presented can matter as much as what is presented.
Tool-augmented prompting strategies such as ReAct~\cite{yao2022react} interleave reasoning and tool calls (e.g., search, open file, run tests), while agent-computer interfaces for repository-level development enable iterative retrieval and refinement~\cite{wang2024openhands}.
Patch-format constraints (e.g., unified diff formats or structured search-and-replace blocks) are frequently adopted to ensure minimal and verifiable edits.

\subsection{Empirical Studies on the Impact of Fault Localization in APR}

Prior empirical research has examined how fault localization quality influences APR effectiveness from several perspectives.

Yang et al.~\cite{yang2017empirical,yang2021evaluating} analyze how APR tools consume localization results, distinguishing between \emph{restrictive} strategies that limit repair to suspicious locations and \emph{prioritized} strategies that explore additional locations when necessary. Their findings demonstrate that repair effectiveness depends not only on localization accuracy but also on how localization information is integrated into the repair workflow.

Similarly, Assiri et al.~\cite{assiri2017fault} show that even moderate inaccuracies in suspiciousness rankings can significantly reduce the likelihood of generating correct patches, highlighting the strong dependency between localization precision and repair capability. Liu et al.~\cite{liu2021critical} further argue that many APR evaluations fail to separate localization errors from patch-generation limitations. 
They advocate evaluating repair systems under perfect (ground-truth) localization to establish upper bounds and quantify the extent to which FL constrains performance. Their analysis shows that localization often constitutes a major bottleneck.

Other studies investigate alternative localization mechanisms designed to reduce irrelevant candidate regions. For example, Guo et al.~\cite{guo2018empirical} evaluate dynamic slicing as a means of narrowing repair search space. While slicing can reduce search space and improve efficiency, the results indicate that its effect on final repair success is inconsistent and bug-dependent.

\subsection{Positioning of Our Work}

Collectively, prior empirical studies establish that: (1) localization quality strongly influences APR outcomes, (2) repair tools differ in their sensitivity to localization precision, (3) evaluating upper bounds under perfect localization is essential for understanding system capability, and (4) Structural techniques such as slicing can help, but may introduce trade-offs.

However, existing work typically evaluates localization at a single granularity, most commonly statement-level ranking, and rarely examines how localization decisions interact across multiple levels of abstraction. In particular, prior studies do not systematically investigate how \emph{file}-, \emph{element}-, and \emph{line}-level context jointly influence repair performance, how expanding beyond ground-truth localization affects outcomes, or how heuristic expansion strategies compare with LLM-based semantic selection.

In contrast, our work conducts a large-scale factorial empirical study that systematically varies localization granularity \emph{(file, element, and line)} and expansion strategy across 61 configurations. This enables us to quantify not only whether fault localization matters, but \emph{how much}, \emph{at which granularity}, and \emph{under which expansion strategies} localization improves or degrades LLM-based repair.

%% file: Sections/8_Threats_to_validity.tex
\section{Threats to Validity}
\label{sec:threats}

Our results may be influenced by the specific prompt formulations and fault-localization implementations used. Although we mitigate this by applying identical prompts and pipelines across all configurations, alternative prompts or localization tools could yield different outcomes.

We evaluate each configuration using a single repair attempt per instance, which does not fully capture the stochastic variability of LLM-based generation. However, our factorial design includes 61 distinct context configurations evaluated on the same benchmark instances. Because each instance is repeatedly exposed to diverse context expansions across \emph{file}-, \emph{element}-, and \emph{line}-level dimensions, the aggregate trends reflect consistent behavior across many independent reasoning conditions rather than a single sampled outcome. This partially mitigates randomness effects, although repeated sampling could further strengthen confidence in absolute resolution rates.

To avoid confounding factors (the fault localization accuracy vs. the impact of context expansion strategy), our study assumes a perfect localization (i.e., no FL prediction is done and ground-truth (GT) is used instead). While an almost perfect \emph{file} level localization is realistic, achieving precise \emph{element}- and \emph{line}-level localization remains more challenging with current LLMs. Therefore, results using GT \emph{elements} and \emph{lines} should be interpreted as optimistic upper bounds, and practical gains through predicting those elements and lines may be smaller, which in turn may impact the findings regarding the context expansions impact on those levels.

In terms of generalizability, our study focuses on GPT-5-mini and the SWE-bench Verified benchmark (500 Python instances), and the observed trends may differ for other models, programming languages, benchmarks, or dataset versions. Finally, we measure repair success solely using test-passing patches, which may not fully capture patch quality, maintainability, or developer intent.

%% file: Sections/9_Conclusion.tex
\section{Conclusion}
\label{sec:conclusion}

We presented a large-scale empirical study examining how the fault-localization context influences LLM-based automated program repair. Across 61 configurations spanning \emph{file}-, \emph{element}-, and \emph{line}-level context on 500 SWE-bench Verified instances using GPT-5-mini, we systematically isolated both the independent and interaction effects of context design.

Our results show that not all localization granularity contributes equally to repair success. \emph{File}-level expansion provides the dominant performance gains by guiding the model toward the related region of the repository, while \emph{element}-level expansion offers conditional benefits that depend strongly on file-context quality. In contrast, expanding \emph{line}-level context frequently harms repair performance by introducing noise and diluting localization signals. Across all dimensions, we consistently observe that context quality outweighs quantity, with LLM-based semantic retrieval generally outperforming structural heuristics. Notably, LLM-based file retrieval not only achieves higher repair performance but also incurs lower cost, requiring fewer files and smaller token budgets than rule-based expansion.

More broadly, our findings challenge the common APR assumption that either increasing localization precision or expanding context monotonically improves repair outcomes. Instead, effective repair arises from complementary context combinations: broader semantic understanding at higher abstraction levels (\emph{files} and \emph{elements}) paired with precise, minimal localization at the \emph{line} level. Qualitative analyses further show that excessive or poorly targeted context can redirect reasoning, encourage over-generalization, or increase patch complexity.

Future work includes extending these analyses across additional models, benchmarks, and programming languages, developing LLM-aware fault localization methods that optimize semantic relevance rather than structural coverage, and exploring adaptive or ensemble repair strategies that dynamically tailor context selection to bug characteristics.